\documentclass[final]{IEEEtran} 

\usepackage{xspace}
\usepackage{float}
\usepackage{pbox}
\usepackage{tabu}
\usepackage{tabularx}
\usepackage{booktabs}
\usepackage{multirow}
\usepackage[table,xcdraw]{xcolor}
\usepackage{array}
\newcolumntype{P}[1]{>{\centering\arraybackslash}p{#1}}
\usepackage{pdfpages}
\usepackage{graphicx}
\usepackage{times}
\usepackage{eurosym}
\usepackage[hyphens]{url}
\usepackage{apacite}
\usepackage{hhline}
\usepackage[indentfirst=false,leftmargin=8pt,rightmargin=8pt]{quoting}
\renewcommand{\citepunct}{,\penalty\citepunctpenalty\,}
\renewcommand{\citedash}{--}% optionally

\begin{document}

\title{Smartwatch Games: Encouraging Privacy-Protective Behaviour in a Longitudinal Study}

\author{
    \IEEEauthorblockN{Meredydd Williams\IEEEauthorrefmark{1}, Jason R C Nurse\IEEEauthorrefmark{2}, Sadie Creese\IEEEauthorrefmark{1}}\\
    \IEEEauthorblockA{\IEEEauthorrefmark{1}Department of Computer Science, University of Oxford, UK\\
    \texttt{firstname.lastname@cs.ox.ac.uk}}\\
    \IEEEauthorblockA{\IEEEauthorrefmark{2}School of Computing, University of Kent, UK\\
    \texttt{j.r.c.nurse@kent.ac.uk}}
}

\maketitle
\begin{abstract}
While the public claim concern for their privacy, they frequently appear to overlook it. This disparity between concern and behaviour is known as the Privacy Paradox. Such issues are particularly prevalent on wearable devices. These products can store personal data, such as text messages and contact details. However, owners rarely use protective features. Educational games can be effective in encouraging changes in behaviour. Therefore, we developed the first privacy game for (Android) Wear OS watches. 10 participants used smartwatches for two months, allowing their high-level settings to be monitored. Five individuals were randomly assigned to our treatment group, and they played a dynamically-customised privacy-themed game. To minimise confounding variables, the other five received the same app but lacking the privacy topic. The treatment group improved their protection, with their usage of screen locks significantly increasing (\textit{p} = 0.043). In contrast, 80\% of the control group continued to never restrict their settings. After the posttest phase, we evaluated behavioural rationale through semi-structured interviews. Privacy concerns became more nuanced in the treatment group, with opinions aligning with behaviour. Actions appeared influenced primarily by three factors: convenience, privacy salience and data sensitivity. This is the first smartwatch game to encourage privacy-protective behaviour.
\end{abstract}


\section{Introduction}
\label{sec:one}

\subsection{Study Motivation}

The public claim to be concerned about privacy, as suggested by a range of polls and surveys \cite{MorarConsulting2016,Pike2017}. However, we frequently exhibit behaviour which places our data at risk \cite{Beresford2012,Felt2012}. This disparity between claimed concern and empirical action is known as the Privacy Paradox \cite{Norberg2007}. The situation often arises through a lack of awareness \cite{Deuker2009}. This poses a particular risk to wearables, which are both novel and unfamiliar \cite{Williams2017}. Smartwatches offer exciting functionality, providing interactive apps and online connectivity. They can also store a variety of personal data, from text messages to contact details \cite{Do2017}. Despite this, users rarely use available settings to protect their privacy \cite{Udoh2016}. This has led to the Privacy Paradox being prevalent in this environment \cite{Williams2017}. 

Previous work has suggested that this issue can be mitigated by increasing awareness \cite{Deuker2009}. Therefore, many studies have sought to educate users on privacy matters \cite{Kelley2009,Helou2012}. Unfortunately, highlighting a problem is often not sufficient to change behaviour \cite{Bada2015}. Since privacy is rarely a primary goal \cite{Hughes-Roberts2015}, individuals might lack the motivation to protect their data. If we hope to incentivise protection, privacy should be aligned with user wants \cite{Dolan2010}. Rather than mandating compliance, we can then highlight the empowering aspects of protection. Serious games embed incentives within interactivity, using positive reinforcement to instil knowledge \cite{Kumar2013}. Such apps have succeeded in phishing training \cite{Sheng2007} and network defence \cite{Irvine2005}. However, privacy games have never been developed for smartwatches. In previous work (\textit{under review}), we constructed and evaluated an online prototype. Through a 504-person survey, we found that protective actions could be encouraged. Therefore, to empirically assess behaviour, we develop and evaluate the first smartwatch privacy game.

\subsection{Background and Related Work}

\textbf{Privacy and awareness.} Privacy is a nebulous topic, encompassing confidentiality, anonymity and autonomy \cite{Solove2008}. However, since we target technological behaviour, we scope our focus to these interactions. Therefore, we consider informational privacy: ``\textit{the interest an individual has in controlling, or at least significantly influencing, the handling of data about themselves}'' \cite{Clarke1999}.

Polls repeatedly suggest that the public care about their privacy \cite{Pike2017,MorarConsulting2016}. A 2017 survey found 84\% of US consumers were worried about their data, with 70\% stating their concerns have increased \cite{Pike2017}. Despite these assertions, we rarely act to protect our information. We ignore permissions \cite{Felt2012}, skim policies \cite{Glanville2018} and settle for lax default settings \cite{Bonneau2010}. This attitude-behaviour gap has been labelled the Privacy Paradox \cite{Norberg2007}.

We define the Privacy Paradox as the ``\textit{discrepancy between the expressed concern and the actual behavior of users}'' \cite{Barth2017}. Due to its popularity, it has been deconstructed in many previous studies. Veltri and Ivchenko \citeyear{Veltri2017} explored the influence of cognitive scarcity. Through an experiment with 969 users, they discovered that fatigue encouraged disclosure. They used this factor to partially justify the Privacy Paradox. Hallam and Zanella \citeyear{Hallam2017} adopted the lens of Construal Level Theory, which studies whether concepts are considered abstract or concrete \cite{Trope2010}. If issues are hypothetical or temporally distant, as often the case with privacy, they are frequently deemed to be abstract. The authors describe how this `psychological distance' tends to lead to the topic lacking salience. As a result, they concluded that concerns have little influence on protective behaviour. Our work also considers the influence of privacy salience. While the above studies discuss the topic, we actively seek to mitigate the Privacy Paradox.

Research suggests that increasing awareness should address the Privacy Paradox. Deuker \citeyear{Deuker2009} found a concern-behaviour disparity existed due to bounded rationality, incomplete information and psychological variables. When describing bounded rationality, he highlighted that ``\textit{users' capabilities in processing information and drawing the right conclusions are restricted by nature}''. Since individuals fail to process the technical details, they tend to underestimate the risks of privacy invasion. Through building awareness, he believed that both bounded rationality and incomplete information could be addressed. P{\"o}tzsch \citeyear{Potzsch2009} saw two solutions to the disparity: align concern to behaviour or behaviour to concern. By highlighting privacy risk, individuals should be more likely to act. Indeed, Bartsch and Dienlin \citeyear{Bartsch2016} found that knowledge can increase the chance of protective action. However, informed individuals must also have the motivation to put that knowledge into practice. While these researchers theorised wise solutions, we evaluate the success of a privacy intervention.

Jackson and Wang \citeyear{Jackson2018} successfully mitigated the Privacy Paradox on mobile phones. They used customised notifications, with charts highlighting the discrepancy between a user's attitude and their app permissions. Attitudes were evaluated through a concern questionnaire at the start of the study. Based on their selected permissions, the system then predicted their degree of privacy risk. Through an online simulation, the authors found that the disparity decreased after these notifications were viewed. This is encouraging, and we adopt personalised challenges within our games. However, the Privacy Paradox was studied on mobile phones at a single point in time. In contrast, we evaluate smartwatch interactions over a two-month period.

\textbf{Privacy behaviour change.} Awareness can highlight the existence of a particular risk. However, this is often insufficient to change privacy behaviour \cite{Bada2015}. Sasse et al. \citeyear{Sasse2007} recommended a three-stage approach: raise awareness, give education and provide training. In this manner, individuals have an opportunity to practice and refine their behaviour. Finally, even if users possess the knowledge, they must be incentivised to act \cite{Bada2015}. Our game, introduced in Section \ref{sec:four}, seeks to implement all these approaches.

We explore privacy-protective behaviour through the lens of Protection Motivation Theory (PMT) \cite{Rogers1983}. This model ``\textit{postulates the three crucial components of a fear appeal to be (a) the magnitude of noxiousness of a depicted event; (b) the probability of that event's occurrence; and (c) the efficacy of a protective response}'' \cite{Rogers1975}. It seeks to deconstruct why individuals do (or do not) use protection \cite{Rogers1983}. Therefore, it appeared relevant to our efforts at behaviour change. It is comprised of two primary components: threat appraisal and coping appraisal. The former is informed by the severity and vulnerability of a risk. The rewards of functionality are also taken into account. For the latter, self-efficacy and response efficacy is considered. This is balanced against the costs of protection. Our games sought to influence these components to encourage privacy.

We also considered the Theory of Reasoned Action\footnote{The Theory of Reasoned Action is based on ``\textit{the proposition that an individual's behavior is determined by the individual's behavioral intention to perform that behavior}'' \cite{chang1998predicting}} \cite{Fishbein1979}, but this model does not recognise constraints on action \cite{Briggs2017}. It is not deemed appropriate for skilled tasks \cite{Liska1984}, and privacy protection appears to require skill. While we investigated the Theory of Planned Behaviour\footnote{The Theory of Planned Behaviour ``\textit{states that the proximal determinant of behaviour is the intention to act. The intention, in turn, is influenced by the attitude towards the behaviour, subjective norm, and perceived behavioural control}'' \cite{hardeman2002application}} \cite{Ajzen1991}, this fails to account for susceptibility or response efficacy \cite{Norman1996}. In contrast, PMT aligns well with privacy and has been recommended for security behaviour change \cite{Briggs2017}. 

In non-wearable environments, education has prompted protection. Albayram et al. \citeyear{Albayram2017} encouraged screen lock usage on smartphones. In their 228-person study, they divided their participants between a treatment group and a control group. The former watched an educational video, whereas the latter did not. Both groups reported their smartphone actions in pretest and posttest. The treatment group reported improved behaviour, suggesting the video was persuasive. Since we evaluate the efficacy of an intervention, we also adopt a pretest-posttest two-group design. However, rather than using smartphone self-reports, we study smartwatches empirically. 

Albayram et al. \citeyear{Albayram2017a} later explored whether videos can encourage the use of Two-Factor Authentication (2FA). Through a 2x2x2 design, they generated and evaluated eight videos. Their content varied on whether risk, self-efficacy and contingency were included. When the first two components were highlighted in the videos, participants were found to adopt 2FA. Both risk and self-efficacy are considered within PMT, and we also use the theory to encourage alterations. However, while Albayram et al. \citeyear{Albayram2017a} used Amazon Mechanical Turk, we analyse participants through a field study. Our in-person approach delivers several advantages. Since our behaviour is empirical rather than self-reported, it should be less prone to falsehood \cite{Fielding2006}. With participants using a real smartwatch in a native environment, our findings should also have external validity. Finally, although our in-person approach limited our sample size, it supported rationale extraction through rich interviews.

`Nudging' has become a popular approach to encourage protection \cite{Wisniewski2016}. Wang et al. \citeyear{Wang2014} augmented Facebook to highlight the audience of a person's posts. Through their six-week trial, they found unintended disclosures were decreased. Although temporarily influential, behaviour can revert when nudges are removed \cite{Bruyneel2016}. This approach differs from techniques within serious games. Nudging modifies the choice architecture to encourage certain decisions. In contrast, serious games seek to instil lessons through education and positive reinforcement \cite{Connolly2012}. Since intrinsic motivation can be highly persuasive \cite{Ryan2000}, the latter approach might prove more persistent.

\textbf{Behaviour change games.} Serious games can be defined as ``\textit{any form of interactive computer-based game software...that has been developed with the intention to be more than entertainment}'' \cite{Ritterfeld2009}. Such tools have been highly successful, often considered more persuasive than direct training \cite{Wouters2013}. In non-smartwatch environments, security has been frequently addressed. Anti-Phishing Phil \cite{Sheng2007} challenged users to identify fraudulent URLs. After playing an aquatic game, players were better able to avoid phishing campaigns. We differ by targeting smartwatches, but adopt similar Learning Science principles. For example, we implement reflection \cite{Donovan1999}, where players contemplate their learning experience. We also include story-based agents by using Non-Player Characters to guide the user through our narrative \cite{Moreno2001}. Finally, we use the conceptual-procedural principle by augmenting high-level information with specific instructions \cite{ Rittle-Johnson2002}. These techniques sought to encourage protective behaviour.

Immaculacy \cite{Suknot2014} is a proposed privacy game, in which the user faces dystopian scenarios. Characters progress through challenges by undertaking privacy-protective actions. This encourages reflection on behaviour, and we adopt a similar approach. Vaidya et al. \citeyear{Vaidya2014} considered interactive techniques to teach privacy. Since privacy is inherently complex, they recommended that scenarios be used. We implement scenario-based challenges, developing the first smartwatch privacy game.

\textbf{Smartwatch behaviour.} Although wearables have existed for decades, smartwatches have gained recent popularity. They can be defined as ``\textit{an electronic wristwatch that is able to perform many of the functions of a smartphone}'' \cite{CollinsEnglishDictionary2017}. The environment differs greatly from other contexts, particularly when concerning the topic of privacy. Internet-of-Things (IoT) products (defined as belonging to a ``\textit{global network interconnecting smart objects by means of extended Internet technologies}'' \cite{Miorandi2012}) have been criticised for lacking usability \cite{Williams2016x}. Smartwatches have small screens and few buttons, with this constraining the use of protective settings \cite{Horcher2015, Benbunan-Fich2017}. The devices are also unfamiliar and therefore their navigation is less likely to be understood \cite{Williams2017}. Furthermore, they can possess highly-sensitive data \cite{Al-Sharrah2018}, while often having great vulnerability \cite{HewlettPackardEnterprise2014}. Due to the novelty and idiosyncrasy of this environment, behavioural studies might uncover new insights.

Pizza et al. \citeyear{Pizza2016} evaluated behaviour for 34 days, with their participants possessing wearable cameras. They found smartwatches were most often used as timepieces, though they also provided notifications. Jeong et al. \citeyear{Jeong2017} undertook a 203-day longitudinal study, collecting data on 50 participants. They analysed wear, but never considered security or privacy. Indeed, there have been no empirical studies on smartwatch privacy. Our analysis offers a rare glimpse into how these settings are used.

Smartwatches are a challenging interface for games, since they possess small screens and few buttons. Casano et al. \citeyear{Casano2016} evaluated an app entitled `Estimate It!', which sought to teach measurement and geometry. The game was ported to a Tizen OS watch, and users were engaged in gameplay. Design requirements and usability guidelines have also been created for this environment \cite{Li2017a, JimenezVargas2016}. However, educational games remain greatly underexplored. 

Before undertaking this research, we developed an online prototype of our smartwatch game (\textit{under review}). Through this app, we evaluated behaviour change and qualitative rationale. To achieve this, we recruited 504 smartwatch users through a crowdsourcing platform. The treatment half played the prototype, which included privacy questions and challenges. The control participants did not interact with an application. In pretest and posttest, we solicited concern and behaviour through an online survey. Whereas the Privacy Paradox was mitigated in the treatment group, control actions failed to change. While this study was encouraging, the game was emulated and behaviour was self-reported. To empirically evaluate the matter, we now construct and evaluate the first smartwatch privacy game.

\section{Method}
\label{sec:three}

\subsection{Recruitment}

\textbf{Sampling process.} 10 Huawei Watch 2 devices were purchased for this study. Since we monitored all participants over the same two-month period (to minimise extraneous variables), the size of our sample was practically limited. Individuals were loaned an expensive device, and as such there was a security risk. Therefore, in compliance with our university's ethical requirements, we recruited from the institution's students. This demographic offers decent external validity, since smartwatch owners tend to be young and educated \cite{Desarnauts2016}.

\textbf{Recruitment.} To participate, individuals were required to fulfil three criteria. They had to be full-time university students, and therefore accountable for their device. They also needed to be 18 years or older, so they could provide informed consent. Finally, they had to possess a modern Android smartphone, to allow their watch to be configured. Of those eligible applicants, we sought to prioritise diversity. Rather than compiling a white British sample, we included a range of nationalities. Privacy is inherently cultural \cite{Alashoor2015a}, with research suggesting Asian societies do less to protect personal data \cite{Huang2016a}. Therefore, we explored whether European students would use greater protection than those from Asian nations. We also selected individuals from a variety of degree specialisms. We felt this would be more-representative of the public than choosing technologists.

To recruit, flyers were affixed to notice boards across the halls of the university. Emails were also sent to mailing list curators, who could forward the messages if they wished. Participants were fully informed of, and consented to, the monitoring of their watch settings. In addition to the privacy-relevant data, we also received approval to track font size, screen brightness and battery level. This disguised the purpose of our study, while also ensuring high ethical standards.

\subsection{Experimental Structure}

\begin{figure}[h!]
    \includegraphics[width=0.48\textwidth]{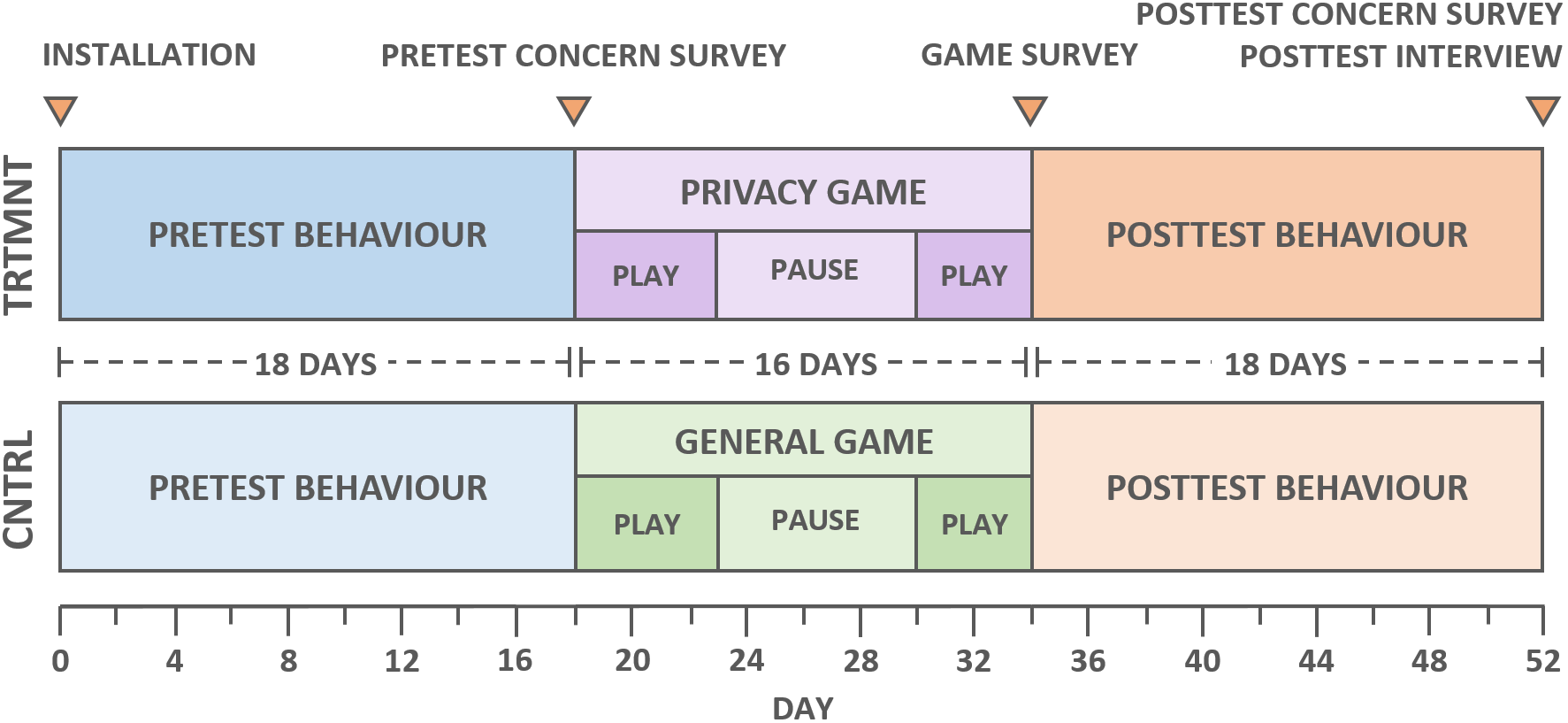}
    \centering
    \caption{Experimental structure}
    \label{fig:structure}
\end{figure}

\textbf{Overview.} Our longitudinal study was divided into three distinct phases: pretest, gameplay and posttest. The experimental structure is shown in Fig \ref{fig:structure}. In a 18-day pretest phase, we monitored the baseline concerns and behaviour of our 10 participants. During a 16-day gameplay phase, these individuals were randomly divided into a treatment group and a control group. Group allocation was truly random, with the process undertaken before study commencement. We considered matching, but thought pure randomisation would reflect the external environment. Fortunately, our groups still appeared well-matched on demographics.

The treatment group (\textit{n} = 5) received a customised privacy game, including challenges to refine behaviour. To minimise confounding variables, the control group (\textit{n} = 5) received the same app, but without the privacy theme. Their game concerned general smartwatch use, such as using gestures and adjusting screen brightness. We originally considered having the control group play no game. However, we were concerned that treatment participants might adjust behaviour purely due to study interaction. Therefore, to reduce bias from the Hawthorne Effect \cite{Adair1984}, both groups received a game. All participants played their game over a five-day and four-day period, with a one-week gap in the middle. Such two-stage approaches can help to construct mental models \cite{Mayer2002}. After these periods were completed, we ended with an 18-day posttest phase. This allowed us to explore whether actions had changed.

We decided against including control variables in our study. We could have considered users' smartwatch familiarity, but all participants lacked prior experience. By collecting non-watch privacy opinions, we might have identified baseline concerns. However, since we sought to disguise the topic (to avoid priming), we decided against this approach.

\textbf{Pretest.} In total, we monitored concerns and behaviour over a 52-day period. We were limited to this span due to the term lengths of our student participants. On day one, 10 individuals were given a Wear OS smartwatch. Once configuration was complete, the monitoring app was installed on each device. This (ethically-approved) service logged settings every five minutes, with details outlined in Subsection \ref{features}. At the end of the phase, we distributed a concern questionnaire to each participant. Its queries can be found in the appendices as Table \ref{tbl:concerns}. We considered soliciting these opinions at the start of the study. However, users might be unfamiliar with smartwatches and therefore unable to provide informed responses. As none of the participants had used such a device before (as revealed in our posttest interviews), our notion was validated. We also sought to assess concern directly before gameplay, since we wished to explore our games' influence. At the end of the pretest phase, individuals had used their smartwatches for 18 days. Therefore, they should have now been able to provide informed opinions.

\textbf{Gameplay.} While users completed the questionnaire, we installed one of two games on their watches. The treatment group received a privacy-themed game, which sought to encourage protection. The control group were given an app with identical gameplay, but concerning a different theme. Rather than the challenges (highlighted below) targeting privacy, they related to general smartwatch usage (e.g., adjusting screen brightness). Since both groups received interactive games, we restricted the influence of extraneous variables. Users were instructed to play the games three times per day for 10 minutes each time. At the end of this 16-day phase, participants completed an evaluation questionnaire. These questions can be found in Table \ref{tbl:evalquestionnaire} of the appendices. This sought to inform future refinements to the games. The questions did not concern privacy, since we did not wish to prime the topic before the posttest phase.

\textbf{Posttest.} To prevent further gameplay influencing behaviour, the apps were uninstalled from the watches. For the final 18 days, we continued to monitor privacy behaviour. At the end, we distributed identical concern questionnaires to pretest. This enabled analysis of whether opinions changed as the study progressed. Since these forms were completed 18 days after gameplay, we doubt concerns were unfairly primed. 

Finally, we conducted semi-structured interviews to explore behavioural rationale. The questions were all open-ended and can be found in Table \ref{tbl:interview} of the appendices. Smartwatches were then reset while users received their compensation. They were each compensated with a \pounds 40 voucher and entry into a \pounds 70 draw. They were also debriefed on our privacy focus, as this was disguised in forms and recruitment. With concerns and behaviour collected, we could now analyse the Privacy Paradox.

\subsection{Threat Model}

As will be described, we evaluated privacy concerns through hypothetical scenarios. For these concerns to be assessed fairly, we must define a reasonable threat model. 

All individuals made use of a Wear OS smartwatch. This watch contains a number of apps, with some developed by companies other than Google/Huawei. Apps are constrained by permissions, allowing data access to be restricted. Data which is read is often shared (potentially anonymised or aggregated) with external parties \cite{Schneier2015}. The watches can access GPS, offering location-customised functionality. While this provides navigational benefits, the device's current location is accessed \cite{Ashbrook2003a}. These watches are also small, expensive, and consumer-oriented. Like Android smartphones, this places them at a reasonable risk of loss or theft \cite{Matthews2016}. Indeed, their ``\textit{size and portability makes them easy to steal}'' \cite{Baggili2015}. Therefore, if a threat is encountered, it would likely come from app companies or petty criminals.

\subsection{Protective Features and Concern Scenarios}
\label{features}

\textbf{Selection.} To gauge privacy concern, we were required to solicit personal opinions. However, privacy is highly contextual \cite{Nissenbaum2009}, and this can challenge a simple rating \cite{Paine2007}. For example, `very concerned' means little when it is divorced from the particular situation. Context is also important when comparing concerns against behaviour. Trepte et al. \citeyear{Trepte2014} were critical of works that juxtaposed abstract opinions against concrete actions. To compare these factors, it is wise to situate them within the same context. We adapted the design of Lee et al.'s influential work \citeyear{Lee2016b}, by requesting responses to hypothetical scenarios. This supports an analysis grounded within the smartwatch environment. To enable a fair evaluation, scenarios were selected through three criteria:

\begin{enumerate}
\item The issue must be feasible and part of our threat model.
\item The situation should be comprehensible to our sample.
\item Most importantly, there should be a direct correspondence between scenario and privacy-protective tools.
\end{enumerate}

Responses were made on a five-point Likert Scale, also adapted from Lee et al. \citeyear{Lee2016b}, which ranged from \textit{Indifferent} to \textit{Very Concerned}. As highlighted earlier, these questions can be found in Table \ref{tbl:concerns} of the appendices. Individuals then provided a qualitative justification for each answer. Since concerns can be inflated when insufficiently considered \cite{Baek2014}, these queries provided a pause for reflection. When analysing behaviour, it was important to consider the available protective tools. In Wear OS environments, three features appeared particularly relevant. These comprised of: \textit{app permissions}, \textit{GPS disabling} and \textit{screen locks}. The settings are outlined below, alongside their respective concern scenario.

\textbf{Permissions.} Apps\footnote{Wear OS apps can be standalone watch applications, and do not require a smartphone equivalent.} provide useful functions to the smartwatch owner. To provide these services, they often access personal data. While this access is legitimate, details are commonly traded with third parties \cite{Schneier2015}. Fortunately, as on smartphones, privacy permissions can restrict access. When applications cannot read details, they cannot share them with partners. To gauge concern, we asked users how they would feel if their data was accessed. We also asked how they would react if data was shared with others. If a person is opposed, they can reduce their risk through permissions.

\textbf{GPS.} GPS can support great functionality, such as navigation and fitness tracking. To provide these features, a satellite geolocates the smartwatch. By its very nature, this allows the position of a device to be monitored \cite{Ashbrook2003a}. If an individual wants to limit this, they can easily disable their GPS. Then, when functionality is required, it can be briefly re-enabled. To evaluate concerns, we asked users how they would feel if their position was monitored. We also asked how they would react if this data was shared with others. If a person fears this, disabling GPS can reduce the risk.

\textbf{Screen locks.} Passcodes are well-known barriers, and have been suggested to deter smartphone theft \cite{ConsumerReports2014a}. Since the Watch OS interface is similar to Android, this deterrent could apply to watches. Smartwatches are small, expensive and popular. As a result, they have been deemed a feasible target for theft \cite{Baggili2015}. Through using a screen lock, personal data is better-protected. To gauge concerns, we asked users how they would feel if their missing device was accessed. We also solicited reactions to their apps being used by a stranger. If users are concerned about physical access, a screen lock is a simple solution.

\subsection{Research Questions}
\label{sec:hypo}

We explore whether the Privacy Paradox can be mitigated through an educational game. To achieve this, we must compare concerns and behaviour in pretest and posttest. Furthermore, we must judge our treatment group results against those of our control group. Therefore, it is crucial that we first define our study metrics.

\textbf{Metrics.} Concerns are evaluated based on reactions to the above scenarios. Since our Likert data is ordinal, it is conventional to avoid means. However, if questions consider the same topic, it is deemed acceptable to aggregate the scores \cite{Norman2010,Carifio2008}. By taking means of these responses, we receive \textit{location scores}, \textit{stranger scores} and \textit{app scores}. The Cronbach alpha values for these question pairs were 0.837, 0.204 and 0.631, respectively \cite{Cronbach1951}. Since the second alpha was particularly low, we report responses to the two `stranger' scenarios (stranger app access and stranger app use) separately.

To evaluate behaviour, we developed metrics to summarise participant activity. For the \textit{GPS score}, we calculated the percentage of recordings (taken every five minutes) in which the feature was enabled. Similarly, for the \textit{lock score}, we analysed the percentage of logs in which a lock was present. When assessing permissions, we chose to consider the context of the application. Some permissions were deemed to be innocuous, such as waking the screen or increasing the volume. We made this judgement by considering the personal data that might be accessed. Two permissions concerned particularly private details: precise location and text message contents. We analysed these elements since such details could support privacy invasions \cite{Creese2012}. The \textit{permission score} comprised the average acceptance percentage of these two permissions.

\textbf{Research questions.} We both monitor 52 days of empirical behaviour and conduct 10 in-depth interviews. Through this quantity of data, we seek to address the following questions:

\begin{enumerate}
\item Do smartwatch users take action to protect their data? If not, this has implications for smartwatch risk and interface design.
\item In smartwatch environments, does a Privacy Paradox appear to be prevalent? If so, users might place their personal data at risk.
\item Can the smartwatch game encourage privacy-protective behaviour? If so, such apps could offer an interactive and low-cost complement to awareness campaigns.
\item What factors influence smartwatch behaviour? If we can understand behavioural rationale, we might be better-placed to design interventions.
\end{enumerate}

\section{Game Design and Rationale}
\label{sec:four}

We now move forward to discuss the design of our two smartwatch games. Most attention will be given to the privacy version, since this sought to encourage protective behaviour. A YouTube video of the game can be found at: \url{https://goo.gl/K7DVfL}.

\subsection{Game Narrative and Mechanics}

\textbf{Overview.} Both games challenged users to navigate across a maze-like map. The privacy version can be found below in Figure \ref{fig:priv}. The player starts at their house and then must traverse four levels to reach the shops. En route, they collect coins to increase their score. When the game ends (or is completed), this score is ranked on a competitive leaderboard. 

During their journey, users encounter two types of Non-Player Character (NPC): `villagers' and `thieves'. Villagers ask functionality questions and reward correct answers with points. For example, a player might be asked, ``\textit{How can I prevent apps accessing contact details?}'', and select the ``\textit{Revoke contacts permission}'' response. 

Thieves block the user's path and trigger functionality challenges. In these challenges, characters must configure a settings menu before their health expires. For example, a player might be tasked to enable a screen lock. Success is rewarded by additional coins, while failure ends the game. 

If the final level is completed on normal difficulty, extra modes are unlocked. All aforementioned components are identical on both games. The only differences were that the privacy app's challenges/questions related to protective features.

\begin{figure}[h!]
    \includegraphics[width=0.48\textwidth]{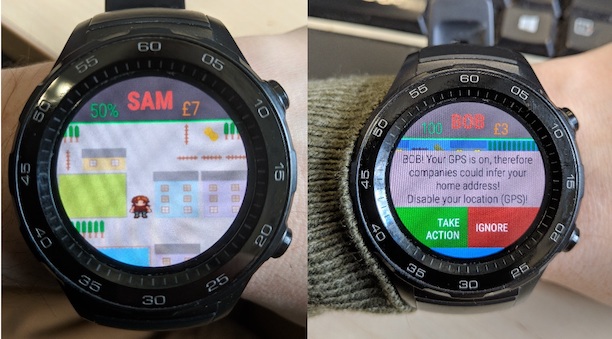}
    \centering
    \caption{Smartwatch privacy game. Left: SAM playing the game in `Morning' mode. Right: BOB facing a customised challenge in `Night' mode.}
    \label{fig:priv}
\end{figure}

\textbf{Challenges.} The general version concerned generic non-privacy smartwatch features. These included adjusting font size, changing screen brightness and configuring alarm volume. In our concern questionnaires, we included decoy questions relating to these settings. Therefore, the purpose of our study was further disguised. The privacy game focused on the three protective approaches: restricting permissions, disabling GPS and enabling a screen lock. These were divided into 14 challenges, shown below in Table \ref{tbl:challenges}.

\begin{table}[h!]
\caption{Privacy Game Challenges: Set tasks in italics.}
\footnotesize
\begin{tabular}{cc} 
\toprule
\multicolumn{2}{c}{Level One (2 Challenges: 1 Set, 1/3 Random)}\\
\midrule
\textit{Disable GPS} & Enable a screen lock pattern\\
Check app permissions & Revoke contacts permissions\\
 \midrule
\multicolumn{2}{c}{Level Two (3 Challenges: 1 Set, 2/3 Remaining)}\\
 \midrule
Enable a screen lock pattern & Check app permissions\\
Revoke contacts permissions & \textit{Revoke audio permissions}\\
\midrule
\multicolumn{2}{c}{Level Three (4 Challenges: 1 Set, 3/3 Random Order)}\\
 \midrule
Enable a screen lock PIN & Revoke location permissions\\
Check system app permissions & \textit{Disable GPS \& location perms}\\
 \midrule
\multicolumn{2}{c}{Level Four (5 Challenges: 5/5 Set Order)}\\
 \midrule
 \textit{Revoke SMS permissions} & \textit{Enable a screen lock password}\\
\textit{Revoke sensors permissions} & \textit{Lock pattern, GPS, location perms}\\
\multicolumn{2}{c}{\textit{Uninstall application}}
\\
\bottomrule
\end{tabular}
\centering
\label{tbl:challenges}
\end{table}

In seeking to highlight each participant's risk exposure, these challenges were dynamically customised around user behaviour. This was achieved through reading the recent log files of the monitoring app. Based on GPS, screen lock, apps installed and app permissions, we contextualised the tasks. For example, a participant might grant ACCESS\_FINE\_LOCATION permissions to their Uber app. If this had occurred, location challenges (in the game) would be customised with these details. This design followed the influential work of Harbach et al. \citeyear{Harbach2014}, who used a similar approach on smartphones. Since our contextualisation was dynamic, the game adjusted to reflect recent user behaviour. This provided an educational feedback loop to encourage protection \cite{Kiili2005}.

\subsection{Behaviour Change Principles}

The games were designed with educational techniques from psychology \cite{Garg2012}, learning science \cite{Quinn2005} and HCI \cite{Richards2014}. They could be defined as `operative games', since they ``\textit{leverage knowledge gained from the study of games or play to exert control upon the world such as encouraging exercise or learning}'' \cite{Carter2014}. 

\textbf{Personalisation.} When participants first open the app, they assign themselves a three-digit name. Since customisation contributes to immersion \cite{Annetta2006}, individuals should continue in a more-retentive manner. They then personalise their character, toggling gender, hair colour and skin colour. These avatars tend to further increase immersion \cite{Annetta2006a}, and might lead to the receptive `flow' state \cite{Kiili2005}.

\begin{figure}[h!]
    \includegraphics[width=0.48\textwidth]{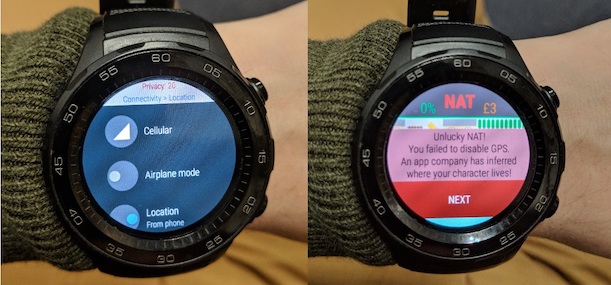}
    \centering
    \caption{Smartwatch privacy challenges. Left: DISABLE\_GPS challenge in progress. Right: DISABLE\_GPS failure screen, describing the consequences.}
    \label{fig:challenge}
\end{figure}

\textbf{Practice.} Game challenges required the configuration of a settings menu. Through completing tasks, participants learned directly-applicable skills. Rather than adjusting the real menu, we implemented a simulated interface (found in Figure \ref{fig:challenge}). This allowed us to provide on-screen hints, found to offer further education \cite{Woolf2010}. It also enabled users to experiment in a safe environment, without being forced to change their own settings. Challenges were time-pressured, encouraging players to remember the menu layout. This sought to trigger `pleasurable frustration', where users enjoy a fun but challenging task \cite{Gee2004}.

\textbf{Education.} Since we wished to enhance privacy knowledge, we included education within the application. When a new game is started, individuals can watch a brief slideshow. The presentation highlights protective approaches and the potential consequences of inaction. In the control group game, the slides concerned general elements of functionality. Again, by including similar features in both apps, we sought to minimise the influence of extraneous variables.

\textbf{Contextualisation.} We selected an accessible real-world narrative: that of going to the shops. Although we considered more exciting scenarios, we took guidance from the literature \cite{Maldonado2005} and selected a relatable situation. Since understanding is enhanced by aligning physical and virtual risks \cite{Garg2012}, we also matched challenges to possible real-world situations. For example, the character faces a challenge when they are near their (gameplay) house. This task requires GPS disabling, since their home location is being `tracked'. In a later level, a thief and villager are adjacent to each other. Since questions could be overheard, the challenge concerns microphone eavesdropping. By relating risks to real situations, participants might consider threats in the future.

\textbf{Principles.} To encourage protective behaviour, we implemented the four learning science principles (goal-oriented, challenging, contextual and interactive) \cite{Quinn2005}. We achieved this through (1) privacy challenges, (2) difficulty modes, (3) dynamic customisation and (4) rich interactivity. The six principles of educational game design were also implemented \cite{Annetta2010}. This was done through avatars (unique identity), rich narrative (immersion), high responsiveness (interactivity), difficulties (increased complexity), challenges (informed teaching), and feedback (instructional).

\textbf{Behaviour change.} The above paragraphs outline our design techniques but they do not specify our behaviour change mechanism. Primarily, we sought to a) appeal to the availability heuristic \cite{Tversky1973} and b) increase user self-efficacy \cite{Bandura1977}. This heuristic describes how ``\textit{a person evaluates the ... probability of events by availability, i.e., by the ease with which relevant instances come to mind}'' \cite{Tversky1973}. By increasing the salience of privacy, we wished to enhance the perceived risk of data infractions. Through self-efficacy, participants should gain confidence to put new skills into action. In concert, this seeks to increase behavioural control. In an attempt to increase salience, we use gameplay feedback, informative questions and interactive challenges. 

Even when individuals recognise the risks, they need knowledge to protect themselves. Through our educational game, we seek to deliver information and provide an opportunity for practice. This should increase individuals' self-efficacy: the confidence that they have in their own expertise. Aligning with Protection Motivation Theory, if users possess self-efficacy and appreciate the risks, they should be more likely to take action \cite{Rogers1983}. We hope this encourages our participants to change their behaviour to a protective state.

\section{Findings and Discussion}
\label{sec:five}

\subsection{Participants and Techniques}
\label{techs}

\textbf{Participants.} 10 participants used smartwatches for two months. Although four came from the UK, we also had individuals from Ireland, Italy, Russia, Mexico, Singapore and the US. While we suspected that concerns and behaviour might differ by culture, this was not found. A larger sample is required to evaluate the influence of this factor. Eight of the users were male, while two were female. Smartwatch users have been disproportionately male and young \cite{NPDConnectedIntelligence2014}, and this trend appears to continue\footnote{www.statista.com/statistics/739398/us-wearable-penetration-by-age/}. Since many also tend to be well-educated \cite{Desarnauts2016}, our sample has some validity. None of our participants had ever used or owned a smartwatch before. This inexperience should limit the influence from prior familiarity. As mature smartwatches are relatively recent\footnote{www.wareable.com/smartwatches/smartwatch-timeline-history-watches}, an inexperienced sample should be externally valid.

\textbf{Quantitative techniques.} Since our sample size was small, we used non-parametric measures for our behavioural comparisons \cite{Siegal1956}. To significance test independent groups, we selected the Mann-Whitney U Test \citeyear{Mann1947}. If the two samples were related, we chose the Wilcoxon Signed-Rank Test \citeyear{Wilcoxon1945}. We required \textit{p} \textless \xspace 0.05 for significance, though its likelihood is limited by our small sample. We used Cohen's \textit{d} for effect sizes, with 0.2, 0.5 and 0.8 representing small, medium and large, respectively \citeyear{Cohen1977}. This metric is less affected by sample size \cite{Cohen1977}, and gives an indication of the game's influence. We use $\bar{x}$ for means, as is standard notation.

We did not undertake significance testing when comparing concerns or opinions. Since these scores only ranged from 1/5 to 5/5, significance was unlikely in a 10-person sample. We also chose not to apply significance testing to our rationale proportions. As these metrics were based on thematic coding, we preferred to use qualitative analyses. 

\textbf{Qualitative techniques.} Through our questionnaires and interviews, we collected a large quantity of qualitative data. This enabled a rare exploration of the privacy rationale of smartwatch behaviour. To ensure a robust evaluation, we undertook best practice through inductive thematic analysis \cite{Ritchie2003}. 

First, all data was formatted in a consistent manner. For our interviews, the researcher undertook verbatim transcription. This approach provides the most detailed account of a discussion \cite{MacLean2004}, further enhancing our validity. We moved on to label recurring topics and concepts. This was undertaken iteratively, seeking to establish consistency between similar replies. Once labelling began to converge, we divided our topics into subtopics. Through this process, we developed conceptual frameworks. These indices then served as our coding frames. Once coding was completed, we selected vivid examples \cite{Braun2006} of rich participant quotes. These are excerpts which we deemed to exemplify a qualitative theme. To select examples, we reviewed those quotes categorised within each topic. If an excerpt was deemed to explain a matter with clarity, it was presented as a vivid example. Through this approach, we aimed to include qualitative description alongside our quantitative findings. These examples are included throughout this section to illustrate user opinions.

\textbf{Validation.} To maximise our validity, we followed four best-practice procedures. Firstly, since we explored rationale through both questionnaires and interviews, we triangulated our findings \cite{Flick2004}. Secondly, our interviews were analysed through multiple coding \cite{Patton1999}. A second researcher, not familiar with with the authors' topic, also coded the transcripts. We analysed consistency by comparing the theme distributions through `proportion agreement' \cite{Morrissey1974}. We selected this method over Cohen's kappa \cite{Cohen1960} for two reasons. Firstly, there were a large number of themes, reducing the risk that matching is due to chance. Secondly, since responses often mentioned multiple themes, kappa is not appropriate \cite{Cohen1960}. The matching accuracy was 83.4\%, suggesting that raters frequently agreed on the categorisation.

Thirdly, we did not seek to hide deviant cases. Where opinions could not be conveniently grouped, distinct themes were retained. Finally, we used respondent validation to verify our understanding \cite{Brink1991}. Each participant was sent their interview transcript and their assigned codes. They were asked to evaluate the accuracy and to suggest refinements. Fortunately, 100\% of the sample agreed with our decisions. Therefore, we believe our findings adequately encapsulate our participants' rationale.

\subsection{Pretest Findings}

Our pretest concern questionnaires were completed 18 days into the study. This ensured that the participants were familiar with their device, but not in possession of training.

\textbf{Opinions.} Before addressing concerns, the questionnaire assessed general opinions. However, we reserve discussion of these to the posttest section. Importantly, we included an instructional manipulation check within the questions \cite{Oppenheimer2009}. In Question 9, participants were asked to indicate their attentiveness by replying `Strongly Disagree'. This query did not serve other purposes and was not related to privacy. Since all individuals answered correctly, the reliability of our responses was deemed enhanced \cite{Oppenheimer2009}.

We then assessed privacy concerns, analysing reactions to violation scenarios. In addition to our described incidents, we included six decoy questions. Through this technique, privacy priming should have been mitigated. The proportion of those at least `concerned' is illustrated below in Table \ref{tbl:pretest}, while the questions can be found in Table \ref{tbl:concerns} of the appendices.

\begin{table}[h!]
\centering
\caption{Pretest privacy concerns}
\label{tbl:pretest}
\bgroup
\def\arraystretch{1.3}
\begin{tabular}{|c|l|c|}
\hline
\textbf{Q} & \multicolumn{1}{c|}{\textbf{Concern Scenario}} & \textbf{Concerned (\%)}    \\ \hline
16         & GPS: Location tracking         & \cellcolor[HTML]{FFFECD}50 \\ \hline
25         & GPS: Location sharing          & \cellcolor[HTML]{FFECCD}80 \\ \hline
19         & Screen locks: Stranger access           & \cellcolor[HTML]{FFECCD}80 \\ \hline
21         & Screen locks: Stranger app usage        & \cellcolor[HTML]{FFF4CD}70 \\ \hline
18         & Permissions: App data collection       & \cellcolor[HTML]{FCE0D6}90 \\ \hline
23         & Permissions: App data sharing          & \cellcolor[HTML]{FFECCD}80 \\ \hline
\end{tabular}
\egroup
\end{table}

\textbf{GPS.} Our participants expressed some opposition to location tracking, with 50\% indicating their concern. However, opinions varied, with another 30\% being quite indifferent. When assessing these scenarios, we also considered the participants' rationale. 16 justifications were given, and while 37.5\% expressed concerns, 37.5\% were dependent on the situation. For example, 18.8\% claimed their reaction depended on whether the tracking was optional. If the tracking could be disabled, as GPS can be, they would be less worried. However, for location data to be protected, intentions must turn into action. 

To directly illustrate participant opinions, we display vivid examples below \cite{Braun2006}. We also report the participant ID of the cited individual. Users \textit{A}-\textit{E} were in the treatment group, while \textit{F}-\textit{J} were control participants. At this pretest stage, group membership was irrelevant.

\begin{quoting}
``\textit{I want to be able to decide when I can be tracked}'' (\#C).
\end{quoting}

The second concern scenario, considering location sharing, faced strong opposition. For this incident, 80\% were concerned and nobody expressed indifference. We then considered qualitative justifications, with 21 comments provided. The vast majority expressed concern (85.7\%), with most individuals objecting to the principle (23.8\%). Many also thought this was illegal (14.3\%), but it might be consented through privacy policies. If individuals are truly opposed to this sharing, access can be limited by disabling GPS.

\begin{quoting}
``\textit{I would very much feel as though my privacy was invaded}'' (\#A).
\end{quoting}

Whereas concerns were reported, behaviour was collected in our smartwatch logs. At this pretest stage, all 10 users had their GPS enabled. Indeed, not a single person had adjusted this setting. This implies that their location was accessible for the first 18 days. When considering \textbf{RQ1}, it appears that protective actions are rarely taken. Since the public are unlikely to be receive training, this presents a worrying baseline. We hope that through our game, the issue will gain salience.

\textbf{Screen locks.} When considering unauthorised access, participants were worried. 50\% indicated their strong opposition, while not a single respondent was indifferent. This suggests that users generally reject this intrusion. 15 justifications were given, with 60\% of these fearing great damage. Concerns were primarily driven by the security risk (20\%) and the importance of personal data (20\%). This suggests that our participants place value on their smartwatch. Two individuals were less concerned, with one believing that that their password would protect them. If protective features are enabled, the privacy risk might be mitigated.

\begin{quoting}
``\textit{I don't want a stranger to know my whereabouts}'' (\#G).
\end{quoting}

Participants expressed similar concerns over their apps being used. 70\% were in opposition, with only one person expressing indifference. This suggests that application access is strongly rejected. On this occasion, 14 comments were provided for justification. 64.3\% of these feared great damage, with data access being the most common concern (28.6\%). This is understandable, since apps can contain personal details. Only one participant was unconcerned, and they believed their apps were not sensitive (7.1\%). However, if app usage is feared, a screen lock might be appropriate.

\begin{quoting}
``\textit{They could cause issues through contact and they could gain my details}'' (\#E).
\end{quoting}

Fortunately, four participants had enabled a screen lock. Since games had not yet been played, this suggests the feature is well-known. This might be due to the prevalence of smartphone PINs and patterns. The other six participants had never used a lock. Despite their inaction, they still claimed concern over the scenarios. In these situations, it is likely that the setting was never noticed. Considering \textbf{RQ1}, this implies that protective action is far from constant.

\textbf{Permissions.} We next gauged concern towards an app accessing data. Respondents were strongly in opposition, with 90\% disliking this situation. When considering the justifications, 16 different comments were given. While 25\% depended on particular details, 62.5\% expressed strong concern. Many objected to the access on principle (25\%), whether or not it posed a risk. Individuals might not act to prevent an issue, but oppose it at an ideological level. In such cases, a disparity is often found between concern and behaviour.
 
\begin{quoting}
``\textit{I would want them to respect my privacy}'' (\#E).
\end{quoting}

When considering data sharing, 80\% opposed the incident. Our users appear to reject these practices, despite them being commonly found \cite{Schneier2015}. We then considered the qualitative justifications, with 17 comments provided. 58.8\% of these expressed concern, compared to only 17.6\% with little worries. As before, the most popular objection was purely on principle (23.5\%). These individuals found this data sharing to be invasive. If they wish to limit the content, they could choose to change their permissions.

\begin{quoting}
``\textit{I value my online privacy}'' (\#F).
\end{quoting}

To assess empirical behaviour, we analysed the pretest logs. Throughout this 18-day period, not a single participant had restricted their permissions. In fact, these settings had been loosened by 9/10 users. Furthermore, two installed additional apps with sensitive permissions. Since the public are unlikely to receive education, this presents a worrying baseline. 

\textbf{RQ1.} In our first research question, we explored whether smartwatch users protect their data. Although screen locks were used by some, 60\% neglected the feature. GPS was used constantly by all individuals, enabling locations to be identified. In the case of permissions, settings were loosened rather than tightened. Based on these results, smartwatch users rarely behave in a protective manner.

\textbf{RQ2.} Our second research question explored whether the Privacy Paradox was common in this environment. To illustrate the degree of concern, the mean scores are displayed below in Figure \ref{fig:pretest}.

\begin{figure}[h!]
    \includegraphics[width=0.48\textwidth]{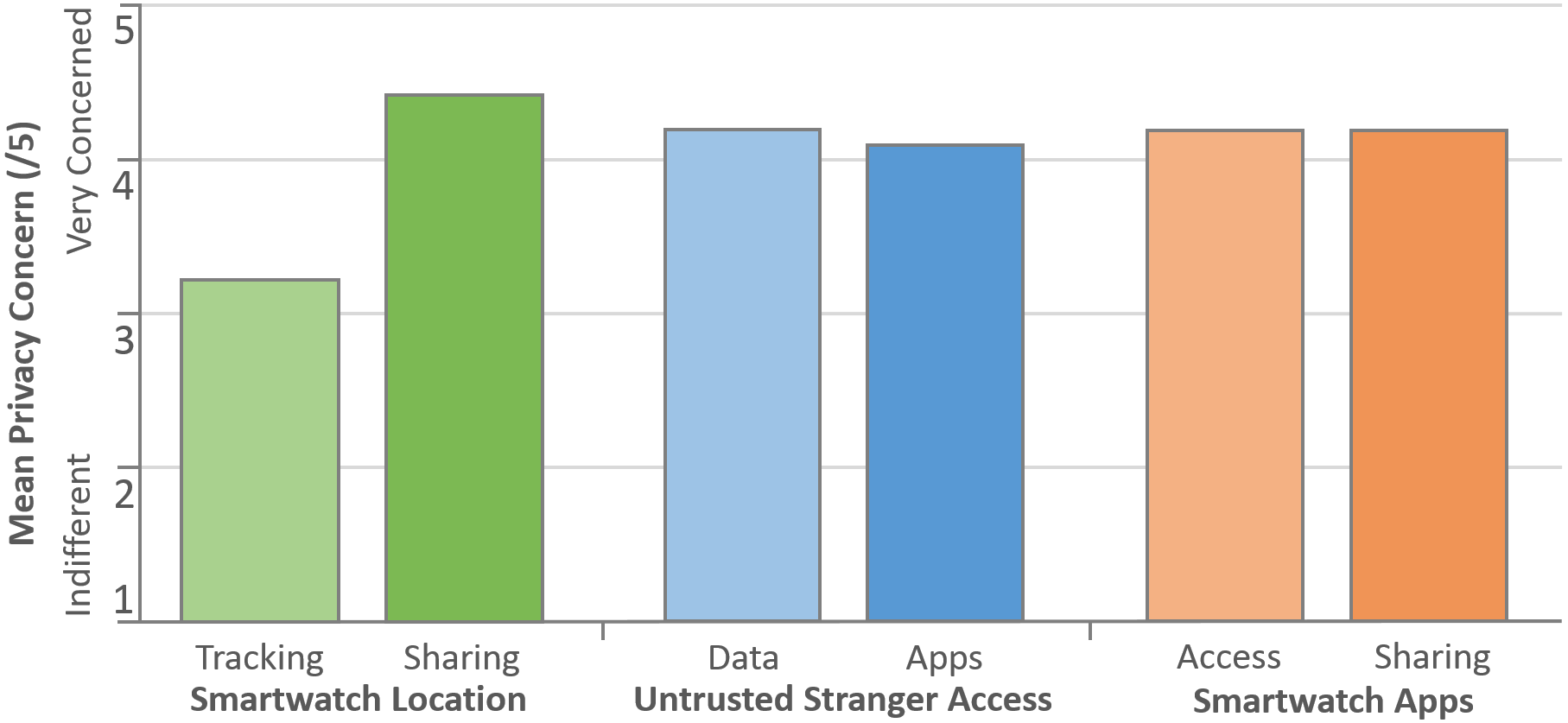}
    \centering
    \caption{Pretest mean privacy concerns.}
    \label{fig:pretest}
\end{figure}

Although location tracking was not strongly opposed, sharing provoked negative responses. This did not encourage any participants to disable their GPS. Our respondents also rejected both unauthorised access and app usage. While some mitigated the risk through passwords, others expressed similar concern. Finally, users appeared to strongly disagree with apps accessing or sharing data. However, they chose to loosen their permissions. Based on these results, concerns and behaviour appear misaligned. Our smartwatch game seeks to mitigate this issue.

\subsection{Gameplay Opinions}

\textbf{Evaluations.} At the end of the gameplay phase, participants completed an evaluation questionnaire. This form can be found in Table \ref{tbl:evalquestionnaire} within the appendices. It did not concern the topic of privacy, since we did not wish to prime posttest behaviour. We first solicited agreement with statements through Likert Scale questions. Users were asked whether they assessed the games as enjoyable, usable, educational and challenging, respectively. Across our sample, 60\% expressed that they enjoyed the games. While it is encouraging that most participants were pleased, we would hope to increase this percentage in future iterations. The agreement level was equal in each group (60\%), suggesting the privacy theme did not detract from enjoyment. 70\% deemed the apps to be usable; promising since usability can encourage retention \cite{Annetta2010}. In this case, 80\% agreed in the control group, compared to 60\% in the treatment group. As will be outlined in Section IV-F, some of the privacy menus were challenging to navigate. This might have contributed to the decreased percentage.

Since we wished to inform participants, we hoped the apps were considered educational. Fortunately, all respondents agreed with the statement. However, differences existed between our groups. Whereas 80\% of control participants were in strong agreement, this was matched by 20\% of the treatment group. Surprisingly, this suggests that the generic game was considered more educational. This opinion might have emerged for two reasons. Firstly, since the privacy app only concentrated on the three protective features, its content was narrower. Secondly, as evidenced in our final Likert-Scale question, privacy tasks were found more challenging. Whereas generic tasks were deemed simple (100\%), nobody thought the same of treatment challenges (0\%). This suggests that privacy is a more-complex topic, and might explain why protective settings are frequently overlooked.

\textbf{Opinions.} We then proceeded to extract opinions through qualitative questions. Their open-ended responses were coded through the thematic analysis highlighted in Subsection \ref{techs}. Firstly, we asked individuals what they most liked about the game. Usability was most praised, with this contributing to 41.2\% of responses. The ease of interaction was particularly appreciated (23.5\%), suggesting our game was simple to play. This is encouraging, since usability has been found to encourage retention \cite{Annetta2010}. When discussing dislikes, participants mentioned 14 factors. The most-frequent complaint was that the games were repetitive (28.6\%). This was partially intentional, since repetition is a standard approach to ingrain knowledge \cite{Franzwa2013}. In future implementations, greater randomness might make the issue less apparent. 

When we asked for suggested improvements, 19 comments were submitted. 57.9\% were in favour of extending the game, with 26.3\% requesting additional `challenges'. This might suggest that users saw feasibility in our approach. While our challenges currently concern installed programs, we could add tasks based on app installation. Through these participant responses, we can refine our games for future interaction.

\subsection{Posttest Concerns}

Users returned their smartwatches at the end of the posttest phase. They then completed final surveys, identical in design to the pretest forms. This allowed fair examination of whether opinions had changed. Since the games had not been played for almost three weeks, they should not prime privacy. Furthermore, our purpose should be disguised by the decoy questions. Due to the small sample sizes, we do not include \textit{p}-values in our below discussion. For ordinal comparisons in a 10-person sample, significance is highly unlikely. As an overview, the pretest-posttest concern proportions (the percentage of those at least responding `concerned') are presented below in Table \ref{tbl:prepost}. The final columns highlight the pretest-posttest change.

\begin{table}[h!]
\centering
\caption[]{Pretest-posttest privacy concerns\\Tmt: Treatment Group, Cnt: Control Group}
\label{tbl:prepost}
\bgroup
\def\arraystretch{1.3}
\begin{tabular}{|l|c|c|c|c|c|c|}
\hline
\multicolumn{1}{|c|}{}                                            & \multicolumn{2}{c|}{\textbf{Pretest (\%)}} & \multicolumn{2}{c|}{\textbf{Posttest (\%)}} & \multicolumn{2}{c|}{\textbf{Change (\%)}}                 \\ \cline{2-7} 
\multicolumn{1}{|c|}{\multirow{-2}{*}{\textbf{Concern Scenario}}} & \textbf{Tmt}         & \textbf{Cnt}        & \textbf{Tmt}         & \textbf{Cnt}         & \textbf{Tmt}                & \textbf{Cnt}                \\ \hline
Location tracking                                                 & 20                   & 80                  & 60                   & 60                   & \cellcolor[HTML]{DAF0DD}+40 & \cellcolor[HTML]{FFECCD}-20 \\ \hline
Location sharing                                                  & 60                   & 100                 & 40                   & 100                  & \cellcolor[HTML]{FFECCD}-20 & \cellcolor[HTML]{FFFECD}0   \\ \hline
Stranger access                                                   & 60                   & 100                 & 100                  & 60                   & \cellcolor[HTML]{DAF0DD}+40 & \cellcolor[HTML]{FCE4D6}-40 \\ \hline
Stranger app usage                                                & 100                  & 40                  & 80                   & 80                   & \cellcolor[HTML]{FFECCD}-20 & \cellcolor[HTML]{DAF0DD}+40 \\ \hline
App data collection                                               & 100                  & 100                 & 60                   & 100                  & \cellcolor[HTML]{FCE4D6}-40 & \cellcolor[HTML]{FFFECD}0   \\ \hline
App data sharing                                                  & 80                   & 80                  & 60                   & 80                   & \cellcolor[HTML]{FFECCD}-20 & \cellcolor[HTML]{FFFECD}0   \\ \hline
\end{tabular}
\egroup
\end{table}

\textbf{Opinions.} In this section, we explored our participants' privacy perceptions. To assess awareness, we asked whether personal data could be read by apps. While agreement decreased in the control group (from $\bar{x}$ = 4.8 to 4.6), it increased in the treatment group (from $\bar{x}$ = 4.4 to 4.8). We then solicited their confidence in their own understanding. As expected, treatment users appeared to have greater self-efficacy than the other group ($\bar{x}$ = 4.6 vs 3.8).

In our third question, participants were asked whether an app might threaten smartwatch data. Whereas the treatment group perceived a threat ($\bar{x}$ = 4.8), controls appeared to lack this knowledge ($\bar{x}$ = 3.4). Although our sample size impedes significance, the privacy game might have enhanced understanding. All users again succeeded in the instructional manipulation check. This implies that responses were made in an engaged manner \cite{Oppenheimer2009}.

\textbf{Location.} When considering location tracking, both groups became more worried. Concerns were now slightly less contingent, and instead focused on the principle of violation (3/10 participants). Users also began to consider targeted advertising and the way their data might be used (2/10). Treatment individuals might have learned about specific risks from their game. Again, representative quotes (with participant ID and group) are shown below.

Firstly, we solicited reactions to location tracking. 60\% were now concerned at the issue (up from 50\%), suggesting one individual might have learned the risk. Although responses are more varied than for some incidents, monitoring appears to provoke some unease. 18 justifications were given, with 55.6\% expressing concern. Treatment reactions were now less dependent, with some opposing the incident on principle (20\%). Control participants feared the leak risk (25\%) but cared less if it was optional (25\%). With respondents expressing greater concern, we hope this contributes to GPS disabling.

\begin{quoting}
``\textit{There still runs a risk that there might be data leakage}'', (\#H, Control).
\end{quoting}

We also analysed reactions to location sharing. Concerns appeared to have altered greatly since the pretest stage. The control group were still opposed, with all respondents being `Very Concerned'. Surprisingly, only two treatment participants acted in the same manner (40\%). Indeed, their concern appeared to decrease as the study progressed. To investigate the rationale, we analysed our 15 qualitative responses. In the control group, individuals feared a security risk (33.3\%). One participant was also worried because they felt uninformed (11.1\%). Three responses were indifferent, with all these coming from treatment users. Participant \textit{D} doubted their risk since they disabled GPS. This report was true, and it implies that behaviour aligned with concerns.

\begin{quoting}
``\textit{They can only do so if I have my location turned on, and as I only use this feature occasionally it wouldn't bother me too much}'', (\#B, Treatment).
\end{quoting}

\textbf{Stranger access.} When considering unauthorised access, both groups showed strong opposition. This matched the pretest reaction, suggesting that this incident is still rejected. If so, more screen locks should have been enabled. While concerns were strong, the rationale differed between our groups. For treatment participants, the access to personal data was most troubling (40\%). They also opposed the security risk that these details could pose (20\%). In contrast, several in the control group doubted their sensitivity (28.6\%). If they had played the privacy game, perhaps they would have knowledge of their risk exposure.

\begin{quoting}
``\textit{Not sure they'd get much out of it}'' (\#I, Control).
\end{quoting}

Both groups continued to oppose unauthorised app use. 80\% expressed concern at the scenario, with the distribution of responses being identical. This was greater than the 70\% in pretest, suggesting the risk might have gained salience. While both groups were predominantly concerned, their qualitative rationale differed. Our treatment participants named specific issues, such as impersonation (33.3\%) or identity theft (22.2\%). The control group were more general, and two individuals expressed dependent concerns. Since our privacy game sought to highlight risk, users might have learned of specific threats.

\begin{quoting}
``\textit{Identity theft is my worst fear}'' (\#A, Treatment).
\end{quoting}

\textbf{App access.} Concerns differed more considerably when discussing data collection. All the control group were worried, with 40\% giving strong responses. In contrast, 40\% of treatment participants supplied a neutral reply. Through the 15 justifications, explored what encouraged these views. Control users were worried about data selling (16.7\%) and the risk of leakage (16.7\%). Treatment participants were alone in offering mitigative views. One expressed that data could be collected through other means (11.1\%). While true, permissions provide a rare opportunity to limit access.

\begin{quoting}
``\textit{Companies already have means of getting so not too concerned}'' (\#D, Treatment).
\end{quoting}

For the final scenario, we assessed reactions to data sharing. As in the previous incident, the treatment group appeared to lose concern. While 80\% of control participants were worried, the others appeared less concerned. To explore why, we analysed the 16 qualitative responses. In the control group, targeted advertising was the main issue (37.5\%). For treatment users, reactions were dependent on other factors. Their concerns were nuanced, based on whether data was sensitive (16.7\%) or aggregated (16.7\%). Rather than scaring users, the privacy game might support informed judgements. Hopefully, they also learned how to adjust their permissions.

\begin{quoting}
``\textit{I wouldn't mind if ... it was information that wasn't too specific}'' (\#B, Treatment).
\end{quoting}

\textbf{Summary.} For good reason, we hesitate from judging a small sample. However, treatment concern appeared to decrease in 4/6 cases. Individuals might now have a greater recognition of how they are acting. If protection is used more frequently, concerns and behaviour might realign.

When assessing responses critically, there might be several reasons for this pattern. Firstly, if initial responses were strong, posttest answers could indicate regression to the mean. This might be due to the random responses of an unengaged sample. We doubt randomness was the primary factor, since users were engaged frequently within study elements. Secondly, treatment participants might have deemed decreased concern to be a study objective. Therefore, their answers were influenced by a response bias. However, through decoy questions at all stages, we sought to disguise the purpose of our study. Finally, the `fear of the unknown' might have magnified the pretest concerns. When smartwatches then became familiar, this effect may have decreased. We believe that this factor is most likely to have proved influential. Whereas initial responses might have been vague, posttest concerns were informed by behavioural experience. Therefore, concern-behaviour alignment might still be an outcome.

\subsection{Posttest Behaviour}

For our third research question (\textbf{RQ3}), we explored whether the game could encourage protective behaviour. To assess this, we monitored the smartwatches for 52 days. If activity differs between pretest and posttest, our app might be influential.

A per-participant comparison can be found in Table \ref{fig:prepost}. Increases in protection are highlighted in green, while deteriorations are in red. Table \ref{fig:time} illustrates the mean daily behaviour throughout the study. The columns denote time periods, as shown in Figure \ref{fig:structure}, while the rows denote participants' actions.

\begin{table}[h!]
\centering
\caption{Pretest-posttest difference in protective behaviour}
\label{fig:prepost}
\tabulinesep=1.2mm
\begin{tabu}{|c|c|P{1.7cm}|P{1.7cm}|P{1.7cm}|}
\hline
\multicolumn{2}{|c|}{\textbf{\#}}                                    & \textbf{GPS}                   & \textbf{Screen Lock}             & \textbf{Permissions}             \\ \hline \hline
\multicolumn{1}{|l|}{}                                           & \textbf{A} & -                              & -                                & +1.1\%                          \\  \hhline{|~|-|-|-|-}
\multicolumn{1}{|l|}{}                                           & \textbf{B} & {\cellcolor[HTML]{9ABD83}}-100\% & {\cellcolor[HTML]{9ABD83}}+99.9\%    & \cellcolor[HTML]{BEDDAA}-22\% \\ \hhline{|~|-|-|-|-} 
\multicolumn{1}{|l|}{}                                           & \textbf{C} & -                              & -                                & \cellcolor[HTML]{F2BD99}+21.5\%    \\ \hhline{|~|-|-|-|-} 
\multicolumn{1}{|l|}{}                                           & \textbf{D} & \cellcolor[HTML]{9ABD83}-100\% & \cellcolor[HTML]{BEDDAA}+13.7\% & +2.4\%                           \\ \hhline{|~|-|-|-|-} 
\multicolumn{1}{|l|}{\multirow{-5}{*}{\rotatebox[origin=c]{90}{\textbf{TREATMENT}}}} & \textbf{E} & -                              & \cellcolor[HTML]{9ABD83}+100\%   & -                                \\ \hhline{=====}
                                                                 & \textbf{F} & -                              & -                                & -                                \\ \cline{2-5} 
                                                                 & \textbf{G} & -                              & -                                & -                                \\ \cline{2-5} 
                                                                 & \textbf{H} & -                              & -                                & +0.1\%                             \\ \cline{2-5} 
                                                                 & \textbf{I} & -                              & -                                & +1.1\%                          \\ \cline{2-5} 
\multirow{-5}{*}{\rotatebox[origin=c]{90}{\textbf{CONTROL}}}                         & \textbf{J} & -                              & -                                & +0.4                                \\ \hline
\end{tabu}
\end{table}

\begin{table*}[t]
\centering
\tiny
\caption[caption]{Longitudinal protective behaviour: (G)PS Usage, (S)creen Lock Usage and (P)ermissions Acceptance.\\GPS: $\leq$ 49\% green and $\geq$ 50\% orange. Screen Lock: $\leq$ 49\% orange and $\geq$ 50\% green.\\Permissions: $\leq$ 33\% green, 34\% - 66\% yellow and $\geq$ 67\% orange.}
\label{fig:time}
\tabcolsep=0.15cm
\tabulinesep=0.9mm
\begin{tabu}{|c|l|l|llllllllllllllllll|lllll|lllllll|llll|llllllllllllllllll|}
\hhline{|-|-|-|------------------|-----|-------|----|------------------|}
\multicolumn{3}{|c|}{\textbf{\#}}                                                                   & \multicolumn{18}{c|}{\textbf{PRETEST PERIOD}}                                                                                                                                                                                                                                                                                                                                                                                                                                                                                                                                                                                                                                                                                                                                                                                                                                                                                     & \multicolumn{5}{c|}{\cellcolor[HTML]{C6D9F4}\textbf{GAME}}                                                                                                                                                                                                                  & \multicolumn{7}{c|}{\cellcolor[HTML]{C6D9F4}\textbf{PAUSE}}                                                                                                                                                                                                                                                                                                                         & \multicolumn{4}{c|}{\cellcolor[HTML]{C6D9F4}\textbf{GAME}}                                                                                                                                                                & \multicolumn{18}{c|}{\textbf{POSTTEST PERIOD}}                                                                                                                                                                                                                                                                                                                                                                                                                                                                                                                                                                                                                                                                                            \\ \hhline{=======================================================}
                                            & \multicolumn{1}{c|}{}                             & G & \cellcolor[HTML]{F4B083}{\color[HTML]{FFCE93} } & \cellcolor[HTML]{F4B083}{\color[HTML]{FFCE93} } & \cellcolor[HTML]{F4B083}{\color[HTML]{FFCE93} } & \cellcolor[HTML]{F4B083}{\color[HTML]{FFCE93} } & \cellcolor[HTML]{F4B083}{\color[HTML]{FFCE93} } & \cellcolor[HTML]{F4B083}{\color[HTML]{FFCE93} } & \cellcolor[HTML]{F4B083}{\color[HTML]{FFCE93} } & \cellcolor[HTML]{F4B083}{\color[HTML]{FFCE93} } & \cellcolor[HTML]{F4B083}{\color[HTML]{FFCE93} } & \cellcolor[HTML]{F4B083}{\color[HTML]{FFCE93} } & \cellcolor[HTML]{F4B083}{\color[HTML]{FFCE93} } & \cellcolor[HTML]{F4B083}{\color[HTML]{FFCE93} } & \cellcolor[HTML]{F4B083}{\color[HTML]{FFCE93} } & \cellcolor[HTML]{F4B083}{\color[HTML]{FFCE93} } & \cellcolor[HTML]{F4B083}{\color[HTML]{FFCE93} } & \cellcolor[HTML]{F4B083}{\color[HTML]{FFCE93} } & \cellcolor[HTML]{F4B083}{\color[HTML]{FFCE93} } & \cellcolor[HTML]{F4B083}{\color[HTML]{FFCE93} } & \cellcolor[HTML]{F4B083}{\color[HTML]{FFCE93} } & \cellcolor[HTML]{F4B083}{\color[HTML]{FFCE93} } & \cellcolor[HTML]{F4B083}{\color[HTML]{FFCE93} } & \cellcolor[HTML]{F4B083}{\color[HTML]{FFCE93} } & \cellcolor[HTML]{F4B083}{\color[HTML]{FFCE93} } & \cellcolor[HTML]{F4B083}{\color[HTML]{FFCE93} } & \cellcolor[HTML]{F4B083}{\color[HTML]{FFCE93} } & \cellcolor[HTML]{F4B083}{\color[HTML]{FFCE93} } & \cellcolor[HTML]{F4B083}{\color[HTML]{FFCE93} } & \cellcolor[HTML]{F4B083}{\color[HTML]{FFCE93} } & \cellcolor[HTML]{F4B083}{\color[HTML]{FFCE93} } & \cellcolor[HTML]{F4B083}{\color[HTML]{FFCE93} } & \cellcolor[HTML]{F4B083}{\color[HTML]{FFCE93} } & \cellcolor[HTML]{F4B083}{\color[HTML]{FFCE93} } & \cellcolor[HTML]{F4B083}{\color[HTML]{FFCE93} } & \cellcolor[HTML]{F4B083}{\color[HTML]{FFCE93} } & \cellcolor[HTML]{F4B083}{\color[HTML]{FFCE93} } & \cellcolor[HTML]{F4B083}{\color[HTML]{FFCE93} } & \cellcolor[HTML]{F4B083}{\color[HTML]{FFCE93} } & \cellcolor[HTML]{F4B083}{\color[HTML]{FFCE93} } & \cellcolor[HTML]{F4B083}{\color[HTML]{FFCE93} } & \cellcolor[HTML]{F4B083}{\color[HTML]{FFCE93} } & \cellcolor[HTML]{F4B083}{\color[HTML]{FFCE93} } & \cellcolor[HTML]{F4B083}{\color[HTML]{FFCE93} } & \cellcolor[HTML]{F4B083}{\color[HTML]{FFCE93} } & \cellcolor[HTML]{F4B083}{\color[HTML]{FFCE93} } & \cellcolor[HTML]{F4B083} & \cellcolor[HTML]{F4B083} & \cellcolor[HTML]{F4B083} & \cellcolor[HTML]{F4B083} & \cellcolor[HTML]{F4B083} & \cellcolor[HTML]{F4B083} & \cellcolor[HTML]{F4B083} & \cellcolor[HTML]{F4B083} \\ \hhline{|~|~|-|~~~~~~~~~~~~~~~~~~|~~~~~|~~~~~~~|~~~~|~~~~~~~~~~~~~~~~~~|}
                                            & \multicolumn{1}{c|}{}                             & S & \cellcolor[HTML]{A8D08D}                        & \cellcolor[HTML]{A8D08D}                        & \cellcolor[HTML]{A8D08D}                        & \cellcolor[HTML]{A8D08D}                        & \cellcolor[HTML]{A8D08D}                      & \cellcolor[HTML]{A8D08D}                        & \cellcolor[HTML]{A8D08D}                        & \cellcolor[HTML]{A8D08D}                        & \cellcolor[HTML]{A8D08D}                        & \cellcolor[HTML]{A8D08D}                        & \cellcolor[HTML]{A8D08D}                        & \cellcolor[HTML]{A8D08D}                        & \cellcolor[HTML]{A8D08D}                        & \cellcolor[HTML]{A8D08D}                        & \cellcolor[HTML]{A8D08D}                        & \cellcolor[HTML]{A8D08D}                        & \cellcolor[HTML]{A8D08D}                        & \cellcolor[HTML]{A8D08D}                        & \cellcolor[HTML]{A8D08D}                        & \cellcolor[HTML]{A8D08D}                        & \cellcolor[HTML]{A8D08D}                        & \cellcolor[HTML]{A8D08D}                        & \cellcolor[HTML]{A8D08D}                        & \cellcolor[HTML]{A8D08D}                        & \cellcolor[HTML]{A8D08D}                        & \cellcolor[HTML]{A8D08D}                        & \cellcolor[HTML]{A8D08D}                        & \cellcolor[HTML]{A8D08D}                        & \cellcolor[HTML]{A8D08D}                        & \cellcolor[HTML]{A8D08D}                        & \cellcolor[HTML]{A8D08D}                        & \cellcolor[HTML]{A8D08D}                        & \cellcolor[HTML]{A8D08D}                        & \cellcolor[HTML]{A8D08D}                        & \cellcolor[HTML]{A8D08D}                        & \cellcolor[HTML]{A8D08D}                        & \cellcolor[HTML]{A8D08D}                        & \cellcolor[HTML]{A8D08D}                        & \cellcolor[HTML]{A8D08D}                        & \cellcolor[HTML]{A8D08D}                        & \cellcolor[HTML]{A8D08D}                        & \cellcolor[HTML]{A8D08D}                        & \cellcolor[HTML]{A8D08D}                        & \cellcolor[HTML]{A8D08D}                        & \cellcolor[HTML]{A8D08D} & \cellcolor[HTML]{A8D08D} & \cellcolor[HTML]{A8D08D} & \cellcolor[HTML]{A8D08D} & \cellcolor[HTML]{A8D08D} & \cellcolor[HTML]{A8D08D} & \cellcolor[HTML]{A8D08D} & \cellcolor[HTML]{A8D08D} \\ \hhline{|~|~|-|~~~~~~~~~~~~~~~~~~|~~~~~|~~~~~~~|~~~~|~~~~~~~~~~~~~~~~~~|}
                                            & \multicolumn{1}{c|}{\multirow{-3}{*}{\textbf{A}}} & P & \cellcolor[HTML]{FFD966}                        & \cellcolor[HTML]{FFD966}                        & \cellcolor[HTML]{FFD966}                        & \cellcolor[HTML]{FFD966}                        & \cellcolor[HTML]{FFD966}                        & \cellcolor[HTML]{FFD966}                        & \cellcolor[HTML]{FFD966}                        & \cellcolor[HTML]{FFD966}                        & \cellcolor[HTML]{FFD966}                        & \cellcolor[HTML]{FFD966}                        & \cellcolor[HTML]{FFD966}                        & \cellcolor[HTML]{FFD966}                        & \cellcolor[HTML]{FFD966}                        & \cellcolor[HTML]{FFD966}                        & \cellcolor[HTML]{FFD966}                        & \cellcolor[HTML]{FFD966}                        & \cellcolor[HTML]{FFD966}                        & \cellcolor[HTML]{FFD966}                        & \cellcolor[HTML]{FFD966}                        & \cellcolor[HTML]{FFD966}                        & \cellcolor[HTML]{FFD966}                        & \cellcolor[HTML]{FFD966}                        & \cellcolor[HTML]{FFD966}                        & \cellcolor[HTML]{FFD966}                        & \cellcolor[HTML]{FFD966}                        & \cellcolor[HTML]{FFD966}                        & \cellcolor[HTML]{FFD966}                        & \cellcolor[HTML]{FFD966}                        & \cellcolor[HTML]{FFD966}                        & \cellcolor[HTML]{FFD966}                        & \cellcolor[HTML]{FFD966}                        & \cellcolor[HTML]{FFD966}                        & \cellcolor[HTML]{FFD966}                        & \cellcolor[HTML]{FFD966}                        & \cellcolor[HTML]{FFD966}                        & \cellcolor[HTML]{FFD966}                        & \cellcolor[HTML]{FFD966}                        & \cellcolor[HTML]{FFD966}                        & \cellcolor[HTML]{FFD966}                        & \cellcolor[HTML]{FFD966}                        & \cellcolor[HTML]{FFD966}                        & \cellcolor[HTML]{FFD966}                        & \cellcolor[HTML]{FFD966}                        & \cellcolor[HTML]{FFD966}                        & \cellcolor[HTML]{FFD966} & \cellcolor[HTML]{FFD966} & \cellcolor[HTML]{FFD966} & \cellcolor[HTML]{FFD966} & \cellcolor[HTML]{FFD966} & \cellcolor[HTML]{FFD966} & \cellcolor[HTML]{FFD966} & \cellcolor[HTML]{FFD966} \\ \hhline{|~|-|-|------------------|-----|-------|----|------------------|} 
                                            &                                                   & G & \cellcolor[HTML]{F4B083}                        & \cellcolor[HTML]{F4B083}                        & \cellcolor[HTML]{F4B083}                        & \cellcolor[HTML]{F4B083}                        & \cellcolor[HTML]{F4B083}                        & \cellcolor[HTML]{F4B083}                        & \cellcolor[HTML]{F4B083}                        & \cellcolor[HTML]{F4B083}                        & \cellcolor[HTML]{F4B083}                        & \cellcolor[HTML]{F4B083}                        & \cellcolor[HTML]{F4B083}                        & \cellcolor[HTML]{F4B083}                        & \cellcolor[HTML]{F4B083}                        & \cellcolor[HTML]{F4B083}                        & \cellcolor[HTML]{F4B083}                        & \cellcolor[HTML]{F4B083}                        & \cellcolor[HTML]{F4B083}                        & \cellcolor[HTML]{F4B083}                        & \cellcolor[HTML]{A8D08D}                        & \cellcolor[HTML]{A8D08D}                        & \cellcolor[HTML]{A8D08D}                        & \cellcolor[HTML]{A8D08D}                        & \cellcolor[HTML]{A8D08D}                        & \cellcolor[HTML]{A8D08D}                        & \cellcolor[HTML]{A8D08D}                        & \cellcolor[HTML]{A8D08D}                        & \cellcolor[HTML]{A8D08D}                        & \cellcolor[HTML]{A8D08D}                        & \cellcolor[HTML]{A8D08D}                        & \cellcolor[HTML]{A8D08D}                        & \cellcolor[HTML]{A8D08D}                        & \cellcolor[HTML]{A8D08D}                        & \cellcolor[HTML]{A8D08D}                        & \cellcolor[HTML]{A8D08D}                        & \cellcolor[HTML]{A8D08D}                        & \cellcolor[HTML]{A8D08D}                        & \cellcolor[HTML]{A8D08D}                        & \cellcolor[HTML]{A8D08D}                        & \cellcolor[HTML]{A8D08D}                        & \cellcolor[HTML]{A8D08D}                        & \cellcolor[HTML]{A8D08D}                        & \cellcolor[HTML]{A8D08D}                        & \cellcolor[HTML]{A8D08D}                        & \cellcolor[HTML]{A8D08D}                        & \cellcolor[HTML]{A8D08D} & \cellcolor[HTML]{A8D08D} & \cellcolor[HTML]{A8D08D} & \cellcolor[HTML]{A8D08D} & \cellcolor[HTML]{A8D08D} & \cellcolor[HTML]{A8D08D} & \cellcolor[HTML]{A8D08D} & \cellcolor[HTML]{A8D08D} \\ \hhline{|~|~|-|~~~~~~~~~~~~~~~~~~|~~~~~|~~~~~~~|~~~~|~~~~~~~~~~~~~~~~~~|}
                                            &                                                   & S & \cellcolor[HTML]{F4B083}                        & \cellcolor[HTML]{F4B083}                        & \cellcolor[HTML]{F4B083}                        & \cellcolor[HTML]{F4B083}                        & \cellcolor[HTML]{F4B083}                        & \cellcolor[HTML]{F4B083}                        & \cellcolor[HTML]{F4B083}                        & \cellcolor[HTML]{F4B083}                        & \cellcolor[HTML]{F4B083}                        & \cellcolor[HTML]{F4B083}                        & \cellcolor[HTML]{F4B083}                        & \cellcolor[HTML]{F4B083}                        & \cellcolor[HTML]{F4B083}                        & \cellcolor[HTML]{F4B083}                        & \cellcolor[HTML]{F4B083}                        & \cellcolor[HTML]{F4B083}                        & \cellcolor[HTML]{F4B083}                        & \cellcolor[HTML]{F4B083}                        & \cellcolor[HTML]{A8D08D}                        & \cellcolor[HTML]{A8D08D}                        & \cellcolor[HTML]{A8D08D}                        & \cellcolor[HTML]{A8D08D}                        & \cellcolor[HTML]{A8D08D}                        & \cellcolor[HTML]{A8D08D}                        & \cellcolor[HTML]{A8D08D}                        & \cellcolor[HTML]{A8D08D}                        & \cellcolor[HTML]{A8D08D}                        & \cellcolor[HTML]{A8D08D}                        & \cellcolor[HTML]{A8D08D}                        & \cellcolor[HTML]{A8D08D}                        & \cellcolor[HTML]{A8D08D}                        & \cellcolor[HTML]{A8D08D}                        & \cellcolor[HTML]{A8D08D}                        & \cellcolor[HTML]{A8D08D}                        & \cellcolor[HTML]{A8D08D}                        & \cellcolor[HTML]{A8D08D}                        & \cellcolor[HTML]{A8D08D}                        & \cellcolor[HTML]{A8D08D}                        & \cellcolor[HTML]{A8D08D}                        & \cellcolor[HTML]{A8D08D}                        & \cellcolor[HTML]{A8D08D}                        & \cellcolor[HTML]{A8D08D}                        & \cellcolor[HTML]{A8D08D}                        & \cellcolor[HTML]{A8D08D}                        & \cellcolor[HTML]{A8D08D} & \cellcolor[HTML]{A8D08D} & \cellcolor[HTML]{A8D08D} & \cellcolor[HTML]{A8D08D} & \cellcolor[HTML]{A8D08D} & \cellcolor[HTML]{A8D08D} & \cellcolor[HTML]{A8D08D} & \cellcolor[HTML]{A8D08D} \\ \hhline{|~|~|-|~~~~~~~~~~~~~~~~~~|~~~~~|~~~~~~~|~~~~|~~~~~~~~~~~~~~~~~~|}
                                            & \multirow{-3}{*}{\textbf{B}}                      & P & \cellcolor[HTML]{FFD966}                        & \cellcolor[HTML]{FFD966}                        & \cellcolor[HTML]{FFD966}                        & \cellcolor[HTML]{FFD966}                        & \cellcolor[HTML]{FFD966}                        & \cellcolor[HTML]{FFD966}                        & \cellcolor[HTML]{FFD966}                        & \cellcolor[HTML]{FFD966}                        & \cellcolor[HTML]{FFD966}                        & \cellcolor[HTML]{FFD966}                        & \cellcolor[HTML]{FFD966}                        & \cellcolor[HTML]{FFD966}                        & \cellcolor[HTML]{FFD966}                        & \cellcolor[HTML]{FFD966}                        & \cellcolor[HTML]{FFD966}                        & \cellcolor[HTML]{FFD966}                        & \cellcolor[HTML]{FFD966}                        & \cellcolor[HTML]{FFD966}                        & \cellcolor[HTML]{FFD966}                        & \cellcolor[HTML]{FFD966}                        & \cellcolor[HTML]{FFD966}                        & \cellcolor[HTML]{FFD966}                        & \cellcolor[HTML]{FFD966}                        & \cellcolor[HTML]{FFD966}                        & \cellcolor[HTML]{FFD966}                        & \cellcolor[HTML]{FFD966}                        & \cellcolor[HTML]{FFD966}                        & \cellcolor[HTML]{FFD966}                        & \cellcolor[HTML]{FFD966}                        & \cellcolor[HTML]{FFD966}                        & \cellcolor[HTML]{FFD966}                        & \cellcolor[HTML]{FFD966}                        & \cellcolor[HTML]{FFD966}                        & \cellcolor[HTML]{FFD966}                        & \cellcolor[HTML]{FFD966}                        & \cellcolor[HTML]{FFD966}                        & \cellcolor[HTML]{FFD966}                        & \cellcolor[HTML]{A8D08D}                        & \cellcolor[HTML]{A8D08D}                        & \cellcolor[HTML]{A8D08D}                        & \cellcolor[HTML]{A8D08D}                        & \cellcolor[HTML]{A8D08D}                        & \cellcolor[HTML]{A8D08D}                        & \cellcolor[HTML]{A8D08D}                        & \cellcolor[HTML]{A8D08D} & \cellcolor[HTML]{A8D08D} & \cellcolor[HTML]{A8D08D} & \cellcolor[HTML]{A8D08D} & \cellcolor[HTML]{A8D08D} & \cellcolor[HTML]{A8D08D} & \cellcolor[HTML]{A8D08D} & \cellcolor[HTML]{A8D08D} \\ \hhline{|~|-|-|------------------|-----|-------|----|------------------|} 
                                            &                                                   & G & \cellcolor[HTML]{F4B083}                        & \cellcolor[HTML]{F4B083}                        & \cellcolor[HTML]{F4B083}                        & \cellcolor[HTML]{F4B083}                        & \cellcolor[HTML]{F4B083}                        & \cellcolor[HTML]{F4B083}                        & \cellcolor[HTML]{F4B083}                        & \cellcolor[HTML]{F4B083}                        & \cellcolor[HTML]{F4B083}                        & \cellcolor[HTML]{F4B083}                        & \cellcolor[HTML]{F4B083}                        & \cellcolor[HTML]{F4B083}                        & \cellcolor[HTML]{F4B083}                        & \cellcolor[HTML]{F4B083}                        & \cellcolor[HTML]{F4B083}                        & \cellcolor[HTML]{F4B083}                        & \cellcolor[HTML]{F4B083}                        & \cellcolor[HTML]{F4B083}                        & \cellcolor[HTML]{F4B083}                        & \cellcolor[HTML]{F4B083}                        & \cellcolor[HTML]{F4B083}                        & \cellcolor[HTML]{F4B083}                        & \cellcolor[HTML]{F4B083}                        & \cellcolor[HTML]{F4B083}                        & \cellcolor[HTML]{F4B083}                        & \cellcolor[HTML]{F4B083}                        & \cellcolor[HTML]{F4B083}                        & \cellcolor[HTML]{F4B083}                        & \cellcolor[HTML]{F4B083}                        & \cellcolor[HTML]{F4B083}                        & \cellcolor[HTML]{F4B083}                        & \cellcolor[HTML]{F4B083}                        & \cellcolor[HTML]{F4B083}                        & \cellcolor[HTML]{F4B083}                        & \cellcolor[HTML]{F4B083}                        & \cellcolor[HTML]{F4B083}                        & \cellcolor[HTML]{F4B083}                        & \cellcolor[HTML]{F4B083}                        & \cellcolor[HTML]{F4B083}                        & \cellcolor[HTML]{F4B083}                        & \cellcolor[HTML]{F4B083}                        & \cellcolor[HTML]{F4B083}                        & \cellcolor[HTML]{F4B083}                        & \cellcolor[HTML]{F4B083}                        & \cellcolor[HTML]{F4B083} & \cellcolor[HTML]{F4B083} & \cellcolor[HTML]{F4B083} & \cellcolor[HTML]{F4B083} & \cellcolor[HTML]{F4B083} & \cellcolor[HTML]{F4B083} & \cellcolor[HTML]{F4B083} & \cellcolor[HTML]{F4B083} \\ \hhline{|~|~|-|~~~~~~~~~~~~~~~~~~|~~~~~|~~~~~~~|~~~~|~~~~~~~~~~~~~~~~~~|}
                                            &                                                   & S & \cellcolor[HTML]{A8D08D}                        & \cellcolor[HTML]{A8D08D}                        & \cellcolor[HTML]{A8D08D}                        & \cellcolor[HTML]{A8D08D}                        & \cellcolor[HTML]{A8D08D}                        & \cellcolor[HTML]{A8D08D}                        & \cellcolor[HTML]{A8D08D}                        & \cellcolor[HTML]{A8D08D}                        & \cellcolor[HTML]{A8D08D}                        & \cellcolor[HTML]{A8D08D}                        & \cellcolor[HTML]{A8D08D}                        & \cellcolor[HTML]{A8D08D}                        & \cellcolor[HTML]{A8D08D}                        & \cellcolor[HTML]{A8D08D}                        & \cellcolor[HTML]{A8D08D}                        & \cellcolor[HTML]{A8D08D}                        & \cellcolor[HTML]{A8D08D}                        & \cellcolor[HTML]{A8D08D}                        & \cellcolor[HTML]{A8D08D}                        & \cellcolor[HTML]{A8D08D}                        & \cellcolor[HTML]{A8D08D}                        & \cellcolor[HTML]{A8D08D}                        & \cellcolor[HTML]{A8D08D}                        & \cellcolor[HTML]{A8D08D}                        & \cellcolor[HTML]{A8D08D}                        & \cellcolor[HTML]{A8D08D}                        & \cellcolor[HTML]{A8D08D}                        & \cellcolor[HTML]{A8D08D}                        & \cellcolor[HTML]{A8D08D}                        & \cellcolor[HTML]{A8D08D}                        & \cellcolor[HTML]{A8D08D}                        & \cellcolor[HTML]{A8D08D}                        & \cellcolor[HTML]{A8D08D}                        & \cellcolor[HTML]{A8D08D}                        & \cellcolor[HTML]{A8D08D}                        & \cellcolor[HTML]{A8D08D}                        & \cellcolor[HTML]{A8D08D}                        & \cellcolor[HTML]{A8D08D}                        & \cellcolor[HTML]{A8D08D}                        & \cellcolor[HTML]{A8D08D}                        & \cellcolor[HTML]{A8D08D}                        & \cellcolor[HTML]{A8D08D}                        & \cellcolor[HTML]{A8D08D}                        & \cellcolor[HTML]{A8D08D}                        & \cellcolor[HTML]{A8D08D} & \cellcolor[HTML]{A8D08D} & \cellcolor[HTML]{A8D08D} & \cellcolor[HTML]{A8D08D} & \cellcolor[HTML]{A8D08D} & \cellcolor[HTML]{A8D08D} & \cellcolor[HTML]{A8D08D} & \cellcolor[HTML]{A8D08D} \\ \hhline{|~|~|-|~~~~~~~~~~~~~~~~~~|~~~~~|~~~~~~~|~~~~|~~~~~~~~~~~~~~~~~~|}
                                            & \multirow{-3}{*}{\textbf{C}}                      & P & \cellcolor[HTML]{FFD966}                        & \cellcolor[HTML]{FFD966}                        & \cellcolor[HTML]{FFD966}                        & \cellcolor[HTML]{F4B083}                        & \cellcolor[HTML]{F4B083}                        & \cellcolor[HTML]{F4B083}                        & \cellcolor[HTML]{F4B083}                        & \cellcolor[HTML]{F4B083}                        & \cellcolor[HTML]{F4B083}                        & \cellcolor[HTML]{F4B083}                        & \cellcolor[HTML]{F4B083}                        & \cellcolor[HTML]{F4B083}                        & \cellcolor[HTML]{F4B083}                        & \cellcolor[HTML]{F4B083}                        & \cellcolor[HTML]{F4B083}                        & \cellcolor[HTML]{F4B083}                        & \cellcolor[HTML]{F4B083}                        & \cellcolor[HTML]{F4B083}                        & \cellcolor[HTML]{F4B083}                        & \cellcolor[HTML]{F4B083}                        & \cellcolor[HTML]{F4B083}                        & \cellcolor[HTML]{F4B083}                        & \cellcolor[HTML]{F4B083}                        & \cellcolor[HTML]{F4B083}                        & \cellcolor[HTML]{F4B083}                        & \cellcolor[HTML]{F4B083}                        & \cellcolor[HTML]{F4B083}                        & \cellcolor[HTML]{F4B083}                        & \cellcolor[HTML]{F4B083}                        & \cellcolor[HTML]{F4B083}                        & \cellcolor[HTML]{F4B083}                        & \cellcolor[HTML]{F4B083}                        & \cellcolor[HTML]{F4B083}                        & \cellcolor[HTML]{F4B083}                        & \cellcolor[HTML]{F4B083}                        & \cellcolor[HTML]{F4B083}                        & \cellcolor[HTML]{F4B083}                        & \cellcolor[HTML]{F4B083}                        & \cellcolor[HTML]{F4B083}                        & \cellcolor[HTML]{F4B083}                        & \cellcolor[HTML]{F4B083}                        & \cellcolor[HTML]{F4B083}                        & \cellcolor[HTML]{F4B083}                        & \cellcolor[HTML]{F4B083}                        & \cellcolor[HTML]{F4B083} & \cellcolor[HTML]{F4B083} & \cellcolor[HTML]{F4B083} & \cellcolor[HTML]{F4B083} & \cellcolor[HTML]{F4B083} & \cellcolor[HTML]{F4B083} & \cellcolor[HTML]{F4B083} & \cellcolor[HTML]{F4B083} \\ \hhline{|~|-|-|------------------|-----|-------|----|------------------|} 
                                            &                                                   & G & \cellcolor[HTML]{F4B083}                        & \cellcolor[HTML]{F4B083}                        & \cellcolor[HTML]{F4B083}                        & \cellcolor[HTML]{F4B083}                        & \cellcolor[HTML]{F4B083}                        & \cellcolor[HTML]{F4B083}                        & \cellcolor[HTML]{F4B083}                        & \cellcolor[HTML]{F4B083}                        & \cellcolor[HTML]{F4B083}                        & \cellcolor[HTML]{F4B083}                        & \cellcolor[HTML]{F4B083}                        & \cellcolor[HTML]{F4B083}                        & \cellcolor[HTML]{F4B083}                        & \cellcolor[HTML]{F4B083}                        & \cellcolor[HTML]{F4B083}                        & \cellcolor[HTML]{F4B083}                        & \cellcolor[HTML]{F4B083}                        & \cellcolor[HTML]{F4B083}                        & \cellcolor[HTML]{F4B083}                        & \cellcolor[HTML]{F4B083}                        & \cellcolor[HTML]{F4B083}                        & \cellcolor[HTML]{A8D08D}                        & \cellcolor[HTML]{A8D08D}                        & \cellcolor[HTML]{A8D08D}                        & \cellcolor[HTML]{A8D08D}                        & \cellcolor[HTML]{A8D08D}                        & \cellcolor[HTML]{A8D08D}                        & \cellcolor[HTML]{A8D08D}                        & \cellcolor[HTML]{A8D08D}                        & \cellcolor[HTML]{A8D08D}                        & \cellcolor[HTML]{A8D08D}                        & \cellcolor[HTML]{A8D08D}                        & \cellcolor[HTML]{A8D08D}                        & \cellcolor[HTML]{A8D08D}                        & \cellcolor[HTML]{A8D08D}                        & \cellcolor[HTML]{A8D08D}                        & \cellcolor[HTML]{A8D08D}                        & \cellcolor[HTML]{A8D08D}                        & \cellcolor[HTML]{A8D08D}                        & \cellcolor[HTML]{A8D08D}                        & \cellcolor[HTML]{A8D08D}                        & \cellcolor[HTML]{A8D08D}                        & \cellcolor[HTML]{A8D08D}                        & \cellcolor[HTML]{A8D08D}                        & \cellcolor[HTML]{A8D08D} & \cellcolor[HTML]{A8D08D} & \cellcolor[HTML]{A8D08D} & \cellcolor[HTML]{A8D08D} & \cellcolor[HTML]{A8D08D} & \cellcolor[HTML]{A8D08D} & \cellcolor[HTML]{A8D08D} & \cellcolor[HTML]{A8D08D} \\ \hhline{|~|~|-|~~~~~~~~~~~~~~~~~~|~~~~~|~~~~~~~|~~~~|~~~~~~~~~~~~~~~~~~|}
                                            &                                                   & S & \cellcolor[HTML]{F4B083}                        & \cellcolor[HTML]{F4B083}                        & \cellcolor[HTML]{A8D08D}                        & \cellcolor[HTML]{A8D08D}                        & \cellcolor[HTML]{A8D08D}                        & \cellcolor[HTML]{A8D08D}                        & \cellcolor[HTML]{A8D08D}                        & \cellcolor[HTML]{A8D08D}                        & \cellcolor[HTML]{A8D08D}                        & \cellcolor[HTML]{A8D08D}                        & \cellcolor[HTML]{A8D08D}                        & \cellcolor[HTML]{A8D08D}                        & \cellcolor[HTML]{A8D08D}                        & \cellcolor[HTML]{A8D08D}                        & \cellcolor[HTML]{A8D08D}                        & \cellcolor[HTML]{A8D08D}                        & \cellcolor[HTML]{A8D08D}                        & \cellcolor[HTML]{A8D08D}                        & \cellcolor[HTML]{A8D08D}                        & \cellcolor[HTML]{A8D08D}                        & \cellcolor[HTML]{A8D08D}                        & \cellcolor[HTML]{A8D08D}                        & \cellcolor[HTML]{A8D08D}                        & \cellcolor[HTML]{A8D08D}                        & \cellcolor[HTML]{A8D08D}                        & \cellcolor[HTML]{A8D08D}                        & \cellcolor[HTML]{A8D08D}                        & \cellcolor[HTML]{A8D08D}                        & \cellcolor[HTML]{A8D08D}                        & \cellcolor[HTML]{A8D08D}                        & \cellcolor[HTML]{A8D08D}                        & \cellcolor[HTML]{A8D08D}                        & \cellcolor[HTML]{A8D08D}                        & \cellcolor[HTML]{A8D08D}                        & \cellcolor[HTML]{A8D08D}                        & \cellcolor[HTML]{A8D08D}                        & \cellcolor[HTML]{A8D08D}                        & \cellcolor[HTML]{A8D08D}                        & \cellcolor[HTML]{A8D08D}                        & \cellcolor[HTML]{A8D08D}                        & \cellcolor[HTML]{A8D08D}                        & \cellcolor[HTML]{A8D08D}                        & \cellcolor[HTML]{A8D08D}                        & \cellcolor[HTML]{A8D08D}                        & \cellcolor[HTML]{A8D08D} & \cellcolor[HTML]{A8D08D} & \cellcolor[HTML]{A8D08D} & \cellcolor[HTML]{A8D08D} & \cellcolor[HTML]{A8D08D} & \cellcolor[HTML]{A8D08D} & \cellcolor[HTML]{A8D08D} & \cellcolor[HTML]{A8D08D} \\ \hhline{|~|~|-|~~~~~~~~~~~~~~~~~~|~~~~~|~~~~~~~|~~~~|~~~~~~~~~~~~~~~~~~|}
                                            & \multirow{-3}{*}{\textbf{D}}                      & P & \cellcolor[HTML]{A8D08D}                        & \cellcolor[HTML]{A8D08D}                        & \cellcolor[HTML]{A8D08D}                        & \cellcolor[HTML]{FFD966}                        & \cellcolor[HTML]{FFD966}                        & \cellcolor[HTML]{FFD966}                        & \cellcolor[HTML]{FFD966}                        & \cellcolor[HTML]{FFD966}                        & \cellcolor[HTML]{FFD966}                        & \cellcolor[HTML]{FFD966}                        & \cellcolor[HTML]{FFD966}                        & \cellcolor[HTML]{FFD966}                        & \cellcolor[HTML]{FFD966}                        & \cellcolor[HTML]{FFD966}                        & \cellcolor[HTML]{FFD966}                        & \cellcolor[HTML]{FFD966}                        & \cellcolor[HTML]{FFD966}                        & \cellcolor[HTML]{FFD966}                        & \cellcolor[HTML]{FFD966}                        & \cellcolor[HTML]{FFD966}                        & \cellcolor[HTML]{FFD966}                        & \cellcolor[HTML]{FFD966}                        & \cellcolor[HTML]{FFD966}                        & \cellcolor[HTML]{FFD966}                        & \cellcolor[HTML]{FFD966}                        & \cellcolor[HTML]{FFD966}                        & \cellcolor[HTML]{A8D08D}                        & \cellcolor[HTML]{A8D08D}                        & \cellcolor[HTML]{A8D08D}                        & \cellcolor[HTML]{A8D08D}                        & \cellcolor[HTML]{A8D08D}                        & \cellcolor[HTML]{A8D08D}                        & \cellcolor[HTML]{A8D08D}                        & \cellcolor[HTML]{A8D08D}                        & \cellcolor[HTML]{A8D08D}                        & \cellcolor[HTML]{A8D08D}                        & \cellcolor[HTML]{FFD966}                        & \cellcolor[HTML]{FFD966}                        & \cellcolor[HTML]{FFD966}                        & \cellcolor[HTML]{FFD966}                        & \cellcolor[HTML]{FFD966}                        & \cellcolor[HTML]{FFD966}                        & \cellcolor[HTML]{FFD966}                        & \cellcolor[HTML]{FFD966}                        & \cellcolor[HTML]{FFD966} & \cellcolor[HTML]{FFD966} & \cellcolor[HTML]{FFD966} & \cellcolor[HTML]{FFD966} & \cellcolor[HTML]{FFD966} & \cellcolor[HTML]{FFD966} & \cellcolor[HTML]{FFD966} & \cellcolor[HTML]{FFD966} \\ \hhline{|~|-|-|------------------|-----|-------|----|------------------|} 
                                            &                                                   & G & \cellcolor[HTML]{F4B083}                        & \cellcolor[HTML]{F4B083}                        & \cellcolor[HTML]{F4B083}                        & \cellcolor[HTML]{F4B083}                        & \cellcolor[HTML]{F4B083}                        & \cellcolor[HTML]{F4B083}                        & \cellcolor[HTML]{F4B083}                        & \cellcolor[HTML]{F4B083}                        & \cellcolor[HTML]{F4B083}                        & \cellcolor[HTML]{F4B083}                        & \cellcolor[HTML]{F4B083}                        & \cellcolor[HTML]{F4B083}                        & \cellcolor[HTML]{F4B083}                        & \cellcolor[HTML]{F4B083}                        & \cellcolor[HTML]{F4B083}                        & \cellcolor[HTML]{F4B083}                        & \cellcolor[HTML]{F4B083}                        & \cellcolor[HTML]{F4B083}                        & \cellcolor[HTML]{F4B083}                        & \cellcolor[HTML]{F4B083}                        & \cellcolor[HTML]{F4B083}                        & \cellcolor[HTML]{F4B083}                        & \cellcolor[HTML]{F4B083}                        & \cellcolor[HTML]{F4B083}                        & \cellcolor[HTML]{F4B083}                        & \cellcolor[HTML]{F4B083}                        & \cellcolor[HTML]{F4B083}                        & \cellcolor[HTML]{F4B083}                        & \cellcolor[HTML]{F4B083}                        & \cellcolor[HTML]{F4B083}                        & \cellcolor[HTML]{F4B083}                        & \cellcolor[HTML]{F4B083}                        & \cellcolor[HTML]{F4B083}                        & \cellcolor[HTML]{F4B083}                        & \cellcolor[HTML]{F4B083}                        & \cellcolor[HTML]{F4B083}                        & \cellcolor[HTML]{F4B083}                        & \cellcolor[HTML]{F4B083}                        & \cellcolor[HTML]{F4B083}                        & \cellcolor[HTML]{F4B083}                        & \cellcolor[HTML]{F4B083}                        & \cellcolor[HTML]{F4B083}                        & \cellcolor[HTML]{F4B083}                        & \cellcolor[HTML]{F4B083}                        & \cellcolor[HTML]{F4B083} & \cellcolor[HTML]{F4B083} & \cellcolor[HTML]{F4B083} & \cellcolor[HTML]{F4B083} & \cellcolor[HTML]{F4B083} & \cellcolor[HTML]{F4B083} & \cellcolor[HTML]{F4B083} & \cellcolor[HTML]{F4B083} \\ \hhline{|~|~|-|~~~~~~~~~~~~~~~~~~|~~~~~|~~~~~~~|~~~~|~~~~~~~~~~~~~~~~~~|}
                                            &                                                   & S & \cellcolor[HTML]{F4B083}                        & \cellcolor[HTML]{F4B083}                        & \cellcolor[HTML]{F4B083}                        & \cellcolor[HTML]{F4B083}                        & \cellcolor[HTML]{F4B083}                        & \cellcolor[HTML]{F4B083}                        & \cellcolor[HTML]{F4B083}                        & \cellcolor[HTML]{F4B083}                        & \cellcolor[HTML]{F4B083}                        & \cellcolor[HTML]{F4B083}                        & \cellcolor[HTML]{F4B083}                        & \cellcolor[HTML]{F4B083}                        & \cellcolor[HTML]{F4B083}                        & \cellcolor[HTML]{F4B083}                        & \cellcolor[HTML]{F4B083}                        & \cellcolor[HTML]{F4B083}                        & \cellcolor[HTML]{F4B083}                        & \cellcolor[HTML]{F4B083}                        & \cellcolor[HTML]{A8D08D}                        & \cellcolor[HTML]{A8D08D}                        & \cellcolor[HTML]{A8D08D}                        & \cellcolor[HTML]{A8D08D}                        & \cellcolor[HTML]{A8D08D}                        & \cellcolor[HTML]{A8D08D}                        & \cellcolor[HTML]{A8D08D}                        & \cellcolor[HTML]{A8D08D}                        & \cellcolor[HTML]{A8D08D}                        & \cellcolor[HTML]{A8D08D}                        & \cellcolor[HTML]{A8D08D}                        & \cellcolor[HTML]{A8D08D}                        & \cellcolor[HTML]{A8D08D}                        & \cellcolor[HTML]{A8D08D}                        & \cellcolor[HTML]{A8D08D}                        & \cellcolor[HTML]{A8D08D}                        & \cellcolor[HTML]{A8D08D}                        & \cellcolor[HTML]{A8D08D}                        & \cellcolor[HTML]{A8D08D}                        & \cellcolor[HTML]{A8D08D}                        & \cellcolor[HTML]{A8D08D}                        & \cellcolor[HTML]{A8D08D}                        & \cellcolor[HTML]{A8D08D}                        & \cellcolor[HTML]{A8D08D}                        & \cellcolor[HTML]{A8D08D}                        & \cellcolor[HTML]{A8D08D}                        & \cellcolor[HTML]{A8D08D} & \cellcolor[HTML]{A8D08D} & \cellcolor[HTML]{A8D08D} & \cellcolor[HTML]{A8D08D} & \cellcolor[HTML]{A8D08D} & \cellcolor[HTML]{A8D08D} & \cellcolor[HTML]{A8D08D} & \cellcolor[HTML]{A8D08D} \\ \hhline{|~|~|-|~~~~~~~~~~~~~~~~~~|~~~~~|~~~~~~~|~~~~|~~~~~~~~~~~~~~~~~~|}
\multirow{-15}{*}{\rotatebox[origin=c]{90}{\textbf{TREATMENT GROUP}}} & \multirow{-3}{*}{\textbf{E}}                      & P & \cellcolor[HTML]{FFD966}                        & \cellcolor[HTML]{FFD966}                        & \cellcolor[HTML]{FFD966}                        & \cellcolor[HTML]{FFD966}                        & \cellcolor[HTML]{FFD966}                        & \cellcolor[HTML]{FFD966}                        & \cellcolor[HTML]{FFD966}                        & \cellcolor[HTML]{FFD966}                        & \cellcolor[HTML]{FFD966}                        & \cellcolor[HTML]{FFD966}                        & \cellcolor[HTML]{FFD966}                        & \cellcolor[HTML]{FFD966}                        & \cellcolor[HTML]{FFD966}                        & \cellcolor[HTML]{FFD966}                        & \cellcolor[HTML]{FFD966}                        & \cellcolor[HTML]{FFD966}                        & \cellcolor[HTML]{FFD966}                        & \cellcolor[HTML]{FFD966}                        & \cellcolor[HTML]{FFD966}                        & \cellcolor[HTML]{FFD966}                        & \cellcolor[HTML]{FFD966}                        & \cellcolor[HTML]{FFD966}                        & \cellcolor[HTML]{FFD966}                        & \cellcolor[HTML]{FFD966}                        & \cellcolor[HTML]{FFD966}                        & \cellcolor[HTML]{FFD966}                        & \cellcolor[HTML]{FFD966}                        & \cellcolor[HTML]{FFD966}                        & \cellcolor[HTML]{FFD966}                        & \cellcolor[HTML]{FFD966}                        & \cellcolor[HTML]{FFD966}                        & \cellcolor[HTML]{FFD966}                        & \cellcolor[HTML]{FFD966}                        & \cellcolor[HTML]{FFD966}                        & \cellcolor[HTML]{FFD966}                        & \cellcolor[HTML]{FFD966}                        & \cellcolor[HTML]{FFD966}                        & \cellcolor[HTML]{FFD966}                        & \cellcolor[HTML]{FFD966}                        & \cellcolor[HTML]{FFD966}                        & \cellcolor[HTML]{FFD966}                        & \cellcolor[HTML]{FFD966}                        & \cellcolor[HTML]{FFD966}                        & \cellcolor[HTML]{FFD966}                        & \cellcolor[HTML]{FFD966} & \cellcolor[HTML]{FFD966} & \cellcolor[HTML]{FFD966} & \cellcolor[HTML]{FFD966} & \cellcolor[HTML]{FFD966} & \cellcolor[HTML]{FFD966} & \cellcolor[HTML]{FFD966} & \cellcolor[HTML]{FFD966} \\ \hline \hline
                                            &                                                   & G & \cellcolor[HTML]{F4B083}                        & \cellcolor[HTML]{F4B083}                        & \cellcolor[HTML]{F4B083}                        & \cellcolor[HTML]{F4B083}                        & \cellcolor[HTML]{F4B083}                        & \cellcolor[HTML]{F4B083}                        & \cellcolor[HTML]{F4B083}                        & \cellcolor[HTML]{F4B083}                        & \cellcolor[HTML]{F4B083}                        & \cellcolor[HTML]{F4B083}                        & \cellcolor[HTML]{F4B083}                        & \cellcolor[HTML]{F4B083}                        & \cellcolor[HTML]{F4B083}                        & \cellcolor[HTML]{F4B083}                        & \cellcolor[HTML]{F4B083}                        & \cellcolor[HTML]{F4B083}                        & \cellcolor[HTML]{F4B083}                        & \cellcolor[HTML]{F4B083}                        & \cellcolor[HTML]{F4B083}                        & \cellcolor[HTML]{F4B083}                        & \cellcolor[HTML]{F4B083}                        & \cellcolor[HTML]{F4B083}                        & \cellcolor[HTML]{F4B083}                        & \cellcolor[HTML]{F4B083}                        & \cellcolor[HTML]{F4B083}                        & \cellcolor[HTML]{F4B083}                        & \cellcolor[HTML]{F4B083}                        & \cellcolor[HTML]{F4B083}                        & \cellcolor[HTML]{F4B083}                        & \cellcolor[HTML]{F4B083}                        & \cellcolor[HTML]{F4B083}                        & \cellcolor[HTML]{F4B083}                        & \cellcolor[HTML]{F4B083}                        & \cellcolor[HTML]{F4B083}                        & \cellcolor[HTML]{F4B083}                        & \cellcolor[HTML]{F4B083}                        & \cellcolor[HTML]{F4B083}                        & \cellcolor[HTML]{F4B083}                        & \cellcolor[HTML]{F4B083}                        & \cellcolor[HTML]{F4B083}                        & \cellcolor[HTML]{F4B083}                        & \cellcolor[HTML]{F4B083}                        & \cellcolor[HTML]{F4B083}                        & \cellcolor[HTML]{F4B083}                        & \cellcolor[HTML]{F4B083} & \cellcolor[HTML]{F4B083} & \cellcolor[HTML]{F4B083} & \cellcolor[HTML]{F4B083} & \cellcolor[HTML]{F4B083} & \cellcolor[HTML]{F4B083} & \cellcolor[HTML]{F4B083} & \cellcolor[HTML]{F4B083} \\ \hhline{|~|~|-|~~~~~~~~~~~~~~~~~~|~~~~~|~~~~~~~|~~~~|~~~~~~~~~~~~~~~~~~|}
                                            &                                                   & S & \cellcolor[HTML]{F4B083}                        & \cellcolor[HTML]{F4B083}                        & \cellcolor[HTML]{F4B083}                        & \cellcolor[HTML]{F4B083}                        & \cellcolor[HTML]{F4B083}                        & \cellcolor[HTML]{F4B083}                        & \cellcolor[HTML]{F4B083}                        & \cellcolor[HTML]{F4B083}                        & \cellcolor[HTML]{F4B083}                        & \cellcolor[HTML]{F4B083}                        & \cellcolor[HTML]{F4B083}                        & \cellcolor[HTML]{F4B083}                        & \cellcolor[HTML]{F4B083}                        & \cellcolor[HTML]{F4B083}                        & \cellcolor[HTML]{F4B083}                        & \cellcolor[HTML]{F4B083}                        & \cellcolor[HTML]{F4B083}                        & \cellcolor[HTML]{F4B083}                        & \cellcolor[HTML]{F4B083}                        & \cellcolor[HTML]{F4B083}                        & \cellcolor[HTML]{F4B083}                        & \cellcolor[HTML]{F4B083}                        & \cellcolor[HTML]{F4B083}                        & \cellcolor[HTML]{F4B083}                        & \cellcolor[HTML]{F4B083}                        & \cellcolor[HTML]{F4B083}                        & \cellcolor[HTML]{F4B083}                        & \cellcolor[HTML]{F4B083}                        & \cellcolor[HTML]{F4B083}                        & \cellcolor[HTML]{F4B083}                        & \cellcolor[HTML]{F4B083}                        & \cellcolor[HTML]{F4B083}                        & \cellcolor[HTML]{F4B083}                        & \cellcolor[HTML]{F4B083}                        & \cellcolor[HTML]{F4B083}                        & \cellcolor[HTML]{F4B083}                        & \cellcolor[HTML]{F4B083}                        & \cellcolor[HTML]{F4B083}                        & \cellcolor[HTML]{F4B083}                        & \cellcolor[HTML]{F4B083}                        & \cellcolor[HTML]{F4B083}                        & \cellcolor[HTML]{F4B083}                        & \cellcolor[HTML]{F4B083}                        & \cellcolor[HTML]{F4B083}                        & \cellcolor[HTML]{F4B083} & \cellcolor[HTML]{F4B083} & \cellcolor[HTML]{F4B083} & \cellcolor[HTML]{F4B083} & \cellcolor[HTML]{F4B083} & \cellcolor[HTML]{F4B083} & \cellcolor[HTML]{F4B083} & \cellcolor[HTML]{F4B083} \\ \hhline{|~|~|-|~~~~~~~~~~~~~~~~~~|~~~~~|~~~~~~~|~~~~|~~~~~~~~~~~~~~~~~~|}
                                            & \multirow{-3}{*}{\textbf{F}}                      & P & \cellcolor[HTML]{A8D08D}                        & \cellcolor[HTML]{A8D08D}                        & \cellcolor[HTML]{A8D08D}                        & \cellcolor[HTML]{A8D08D}                        & \cellcolor[HTML]{A8D08D}                        & \cellcolor[HTML]{A8D08D}                        & \cellcolor[HTML]{A8D08D}                        & \cellcolor[HTML]{A8D08D}                        & \cellcolor[HTML]{A8D08D}                        & \cellcolor[HTML]{A8D08D}                        & \cellcolor[HTML]{A8D08D}                        & \cellcolor[HTML]{A8D08D}                        & \cellcolor[HTML]{A8D08D}                        & \cellcolor[HTML]{A8D08D}                        & \cellcolor[HTML]{A8D08D}                        & \cellcolor[HTML]{A8D08D}                        & \cellcolor[HTML]{A8D08D}                        & \cellcolor[HTML]{A8D08D}                        & \cellcolor[HTML]{A8D08D}                        & \cellcolor[HTML]{A8D08D}                        & \cellcolor[HTML]{A8D08D}                        & \cellcolor[HTML]{A8D08D}                        & \cellcolor[HTML]{A8D08D}                        & \cellcolor[HTML]{A8D08D}                        & \cellcolor[HTML]{A8D08D}                        & \cellcolor[HTML]{A8D08D}                        & \cellcolor[HTML]{A8D08D}                        & \cellcolor[HTML]{A8D08D}                        & \cellcolor[HTML]{A8D08D}                        & \cellcolor[HTML]{A8D08D}                        & \cellcolor[HTML]{A8D08D}                        & \cellcolor[HTML]{A8D08D}                        & \cellcolor[HTML]{A8D08D}                        & \cellcolor[HTML]{A8D08D}                        & \cellcolor[HTML]{A8D08D}                        & \cellcolor[HTML]{A8D08D}                        & \cellcolor[HTML]{A8D08D}                        & \cellcolor[HTML]{A8D08D}                        & \cellcolor[HTML]{A8D08D}                        & \cellcolor[HTML]{A8D08D}                        & \cellcolor[HTML]{A8D08D}                        & \cellcolor[HTML]{A8D08D}                        & \cellcolor[HTML]{A8D08D}                        & \cellcolor[HTML]{A8D08D}                        & \cellcolor[HTML]{A8D08D} & \cellcolor[HTML]{A8D08D} & \cellcolor[HTML]{A8D08D} & \cellcolor[HTML]{A8D08D} & \cellcolor[HTML]{A8D08D} & \cellcolor[HTML]{A8D08D} & \cellcolor[HTML]{A8D08D} & \cellcolor[HTML]{A8D08D} \\ \hhline{|~|-|-|------------------|-----|-------|----|------------------|} 
                                            &                                                   & G & \cellcolor[HTML]{F4B083}                        & \cellcolor[HTML]{F4B083}                        & \cellcolor[HTML]{F4B083}                        & \cellcolor[HTML]{F4B083}                        & \cellcolor[HTML]{F4B083}                        & \cellcolor[HTML]{F4B083}                        & \cellcolor[HTML]{F4B083}                        & \cellcolor[HTML]{F4B083}                        & \cellcolor[HTML]{F4B083}                        & \cellcolor[HTML]{F4B083}                        & \cellcolor[HTML]{F4B083}                        & \cellcolor[HTML]{F4B083}                        & \cellcolor[HTML]{F4B083}                        & \cellcolor[HTML]{F4B083}                        & \cellcolor[HTML]{F4B083}                        & \cellcolor[HTML]{F4B083}                        & \cellcolor[HTML]{F4B083}                        & \cellcolor[HTML]{F4B083}                        & \cellcolor[HTML]{F4B083}                        & \cellcolor[HTML]{F4B083}                        & \cellcolor[HTML]{F4B083}                        & \cellcolor[HTML]{F4B083}                        & \cellcolor[HTML]{F4B083}                        & \cellcolor[HTML]{F4B083}                        & \cellcolor[HTML]{F4B083}                        & \cellcolor[HTML]{F4B083}                        & \cellcolor[HTML]{F4B083}                        & \cellcolor[HTML]{F4B083}                        & \cellcolor[HTML]{F4B083}                        & \cellcolor[HTML]{F4B083}                        & \cellcolor[HTML]{F4B083}                        & \cellcolor[HTML]{F4B083}                        & \cellcolor[HTML]{F4B083}                        & \cellcolor[HTML]{F4B083}                        & \cellcolor[HTML]{F4B083}                        & \cellcolor[HTML]{F4B083}                        & \cellcolor[HTML]{F4B083}                        & \cellcolor[HTML]{F4B083}                        & \cellcolor[HTML]{F4B083}                        & \cellcolor[HTML]{F4B083}                        & \cellcolor[HTML]{F4B083}                        & \cellcolor[HTML]{F4B083}                        & \cellcolor[HTML]{F4B083}                        & \cellcolor[HTML]{F4B083}                        & \cellcolor[HTML]{F4B083} & \cellcolor[HTML]{F4B083} & \cellcolor[HTML]{F4B083} & \cellcolor[HTML]{F4B083} & \cellcolor[HTML]{F4B083} & \cellcolor[HTML]{F4B083} & \cellcolor[HTML]{F4B083} & \cellcolor[HTML]{F4B083} \\ \hhline{|~|~|-|~~~~~~~~~~~~~~~~~~|~~~~~|~~~~~~~|~~~~|~~~~~~~~~~~~~~~~~~|}
                                            &                                                   & S & \cellcolor[HTML]{F4B083}                        & \cellcolor[HTML]{F4B083}                        & \cellcolor[HTML]{F4B083}                        & \cellcolor[HTML]{F4B083}                        & \cellcolor[HTML]{F4B083}                        & \cellcolor[HTML]{F4B083}                        & \cellcolor[HTML]{F4B083}                        & \cellcolor[HTML]{F4B083}                        & \cellcolor[HTML]{F4B083}                        & \cellcolor[HTML]{F4B083}                        & \cellcolor[HTML]{F4B083}                        & \cellcolor[HTML]{F4B083}                        & \cellcolor[HTML]{F4B083}                        & \cellcolor[HTML]{F4B083}                        & \cellcolor[HTML]{F4B083}                        & \cellcolor[HTML]{F4B083}                        & \cellcolor[HTML]{F4B083}                        & \cellcolor[HTML]{F4B083}                        & \cellcolor[HTML]{F4B083}                        & \cellcolor[HTML]{F4B083}                        & \cellcolor[HTML]{F4B083}                        & \cellcolor[HTML]{F4B083}                        & \cellcolor[HTML]{F4B083}                        & \cellcolor[HTML]{F4B083}                        & \cellcolor[HTML]{F4B083}                        & \cellcolor[HTML]{F4B083}                        & \cellcolor[HTML]{F4B083}                        & \cellcolor[HTML]{F4B083}                        & \cellcolor[HTML]{F4B083}                        & \cellcolor[HTML]{F4B083}                        & \cellcolor[HTML]{F4B083}                        & \cellcolor[HTML]{F4B083}                        & \cellcolor[HTML]{F4B083}                        & \cellcolor[HTML]{F4B083}                        & \cellcolor[HTML]{F4B083}                        & \cellcolor[HTML]{F4B083}                        & \cellcolor[HTML]{F4B083}                        & \cellcolor[HTML]{F4B083}                        & \cellcolor[HTML]{F4B083}                        & \cellcolor[HTML]{F4B083}                        & \cellcolor[HTML]{F4B083}                        & \cellcolor[HTML]{F4B083}                        & \cellcolor[HTML]{F4B083}                        & \cellcolor[HTML]{F4B083}                        & \cellcolor[HTML]{F4B083} & \cellcolor[HTML]{F4B083} & \cellcolor[HTML]{F4B083} & \cellcolor[HTML]{F4B083} & \cellcolor[HTML]{F4B083} & \cellcolor[HTML]{F4B083} & \cellcolor[HTML]{F4B083} & \cellcolor[HTML]{F4B083} \\ \hhline{|~|~|-|~~~~~~~~~~~~~~~~~~|~~~~~|~~~~~~~|~~~~|~~~~~~~~~~~~~~~~~~|}
                                            & \multirow{-3}{*}{\textbf{G}}                      & P & \cellcolor[HTML]{FFD966}                        & \cellcolor[HTML]{FFD966}                        & \cellcolor[HTML]{FFD966}                        & \cellcolor[HTML]{FFD966}                        & \cellcolor[HTML]{FFD966}                        & \cellcolor[HTML]{FFD966}                        & \cellcolor[HTML]{FFD966}                        & \cellcolor[HTML]{FFD966}                        & \cellcolor[HTML]{FFD966}                        & \cellcolor[HTML]{FFD966}                        & \cellcolor[HTML]{FFD966}                        & \cellcolor[HTML]{FFD966}                        & \cellcolor[HTML]{FFD966}                        & \cellcolor[HTML]{FFD966}                        & \cellcolor[HTML]{FFD966}                        & \cellcolor[HTML]{FFD966}                        & \cellcolor[HTML]{FFD966}                        & \cellcolor[HTML]{FFD966}                        & \cellcolor[HTML]{FFD966}                        & \cellcolor[HTML]{FFD966}                        & \cellcolor[HTML]{FFD966}                        & \cellcolor[HTML]{FFD966}                        & \cellcolor[HTML]{FFD966}                        & \cellcolor[HTML]{FFD966}                        & \cellcolor[HTML]{FFD966}                        & \cellcolor[HTML]{FFD966}                        & \cellcolor[HTML]{FFD966}                        & \cellcolor[HTML]{FFD966}                        & \cellcolor[HTML]{FFD966}                        & \cellcolor[HTML]{FFD966}                        & \cellcolor[HTML]{FFD966}                        & \cellcolor[HTML]{FFD966}                        & \cellcolor[HTML]{FFD966}                        & \cellcolor[HTML]{FFD966}                        & \cellcolor[HTML]{FFD966}                        & \cellcolor[HTML]{FFD966}                        & \cellcolor[HTML]{FFD966}                        & \cellcolor[HTML]{FFD966}                        & \cellcolor[HTML]{FFD966}                        & \cellcolor[HTML]{FFD966}                        & \cellcolor[HTML]{FFD966}                        & \cellcolor[HTML]{FFD966}                        & \cellcolor[HTML]{FFD966}                        & \cellcolor[HTML]{FFD966}                        & \cellcolor[HTML]{FFD966} & \cellcolor[HTML]{FFD966} & \cellcolor[HTML]{FFD966} & \cellcolor[HTML]{FFD966} & \cellcolor[HTML]{FFD966} & \cellcolor[HTML]{FFD966} & \cellcolor[HTML]{FFD966} & \cellcolor[HTML]{FFD966} \\ \hhline{|~|-|-|------------------|-----|-------|----|------------------|} 
                                            &                                                   & G & \cellcolor[HTML]{F4B083}                        & \cellcolor[HTML]{F4B083}                        & \cellcolor[HTML]{F4B083}                        & \cellcolor[HTML]{F4B083}                        & \cellcolor[HTML]{F4B083}                        & \cellcolor[HTML]{F4B083}                        & \cellcolor[HTML]{F4B083}                        & \cellcolor[HTML]{F4B083}                        & \cellcolor[HTML]{F4B083}                        & \cellcolor[HTML]{F4B083}                        & \cellcolor[HTML]{F4B083}                        & \cellcolor[HTML]{F4B083}                        & \cellcolor[HTML]{F4B083}                        & \cellcolor[HTML]{F4B083}                        & \cellcolor[HTML]{F4B083}                        & \cellcolor[HTML]{F4B083}                        & \cellcolor[HTML]{F4B083}                        & \cellcolor[HTML]{F4B083}                        & \cellcolor[HTML]{F4B083}                        & \cellcolor[HTML]{F4B083}                        & \cellcolor[HTML]{F4B083}                        & \cellcolor[HTML]{F4B083}                        & \cellcolor[HTML]{F4B083}                        & \cellcolor[HTML]{F4B083}                        & \cellcolor[HTML]{F4B083}                        & \cellcolor[HTML]{F4B083}                        & \cellcolor[HTML]{F4B083}                        & \cellcolor[HTML]{F4B083}                        & \cellcolor[HTML]{F4B083}                        & \cellcolor[HTML]{F4B083}                        & \cellcolor[HTML]{F4B083}                        & \cellcolor[HTML]{F4B083}                        & \cellcolor[HTML]{F4B083}                        & \cellcolor[HTML]{F4B083}                        & \cellcolor[HTML]{F4B083}                        & \cellcolor[HTML]{F4B083}                        & \cellcolor[HTML]{F4B083}                        & \cellcolor[HTML]{F4B083}                        & \cellcolor[HTML]{F4B083}                        & \cellcolor[HTML]{F4B083}                        & \cellcolor[HTML]{F4B083}                        & \cellcolor[HTML]{F4B083}                        & \cellcolor[HTML]{F4B083}                        & \cellcolor[HTML]{F4B083}                        & \cellcolor[HTML]{F4B083} & \cellcolor[HTML]{F4B083} & \cellcolor[HTML]{F4B083} & \cellcolor[HTML]{F4B083} & \cellcolor[HTML]{F4B083} & \cellcolor[HTML]{F4B083} & \cellcolor[HTML]{F4B083} & \cellcolor[HTML]{F4B083} \\ \hhline{|~|~|-|~~~~~~~~~~~~~~~~~~|~~~~~|~~~~~~~|~~~~|~~~~~~~~~~~~~~~~~~|}
                                            &                                                   & S & \cellcolor[HTML]{F4B083}                        & \cellcolor[HTML]{F4B083}                        & \cellcolor[HTML]{F4B083}                        & \cellcolor[HTML]{F4B083}                        & \cellcolor[HTML]{F4B083}                        & \cellcolor[HTML]{F4B083}                        & \cellcolor[HTML]{F4B083}                        & \cellcolor[HTML]{F4B083}                        & \cellcolor[HTML]{F4B083}                        & \cellcolor[HTML]{F4B083}                        & \cellcolor[HTML]{F4B083}                        & \cellcolor[HTML]{F4B083}                        & \cellcolor[HTML]{F4B083}                        & \cellcolor[HTML]{F4B083}                        & \cellcolor[HTML]{F4B083}                        & \cellcolor[HTML]{F4B083}                        & \cellcolor[HTML]{F4B083}                        & \cellcolor[HTML]{F4B083}                        & \cellcolor[HTML]{F4B083}                        & \cellcolor[HTML]{F4B083}                        & \cellcolor[HTML]{F4B083}                        & \cellcolor[HTML]{F4B083}                        & \cellcolor[HTML]{F4B083}                        & \cellcolor[HTML]{F4B083}                        & \cellcolor[HTML]{F4B083}                        & \cellcolor[HTML]{F4B083}                        & \cellcolor[HTML]{F4B083}                        & \cellcolor[HTML]{F4B083}                        & \cellcolor[HTML]{F4B083}                        & \cellcolor[HTML]{F4B083}                        & \cellcolor[HTML]{F4B083}                        & \cellcolor[HTML]{F4B083}                        & \cellcolor[HTML]{F4B083}                        & \cellcolor[HTML]{F4B083}                        & \cellcolor[HTML]{F4B083}                        & \cellcolor[HTML]{F4B083}                        & \cellcolor[HTML]{F4B083}                        & \cellcolor[HTML]{F4B083}                        & \cellcolor[HTML]{F4B083}                        & \cellcolor[HTML]{F4B083}                        & \cellcolor[HTML]{F4B083}                        & \cellcolor[HTML]{F4B083}                        & \cellcolor[HTML]{F4B083}                        & \cellcolor[HTML]{F4B083}                        & \cellcolor[HTML]{F4B083} & \cellcolor[HTML]{F4B083} & \cellcolor[HTML]{F4B083} & \cellcolor[HTML]{F4B083} & \cellcolor[HTML]{F4B083} & \cellcolor[HTML]{F4B083} & \cellcolor[HTML]{F4B083} & \cellcolor[HTML]{F4B083} \\ \hhline{|~|~|-|~~~~~~~~~~~~~~~~~~|~~~~~|~~~~~~~|~~~~|~~~~~~~~~~~~~~~~~~|}
                                            & \multirow{-3}{*}{\textbf{H}}                      & P & \cellcolor[HTML]{FFD966}                        & \cellcolor[HTML]{FFD966}                        & \cellcolor[HTML]{FFD966}                        & \cellcolor[HTML]{FFD966}                        & \cellcolor[HTML]{FFD966}                        & \cellcolor[HTML]{FFD966}                        & \cellcolor[HTML]{FFD966}                        & \cellcolor[HTML]{FFD966}                        & \cellcolor[HTML]{FFD966}                        & \cellcolor[HTML]{FFD966}                        & \cellcolor[HTML]{FFD966}                        & \cellcolor[HTML]{FFD966}                        & \cellcolor[HTML]{FFD966}                        & \cellcolor[HTML]{FFD966}                        & \cellcolor[HTML]{FFD966}                        & \cellcolor[HTML]{FFD966}                        & \cellcolor[HTML]{FFD966}                        & \cellcolor[HTML]{FFD966}                        & \cellcolor[HTML]{FFD966}                        & \cellcolor[HTML]{FFD966}                        & \cellcolor[HTML]{FFD966}                        & \cellcolor[HTML]{FFD966}                        & \cellcolor[HTML]{FFD966}                        & \cellcolor[HTML]{FFD966}                        & \cellcolor[HTML]{FFD966}                        & \cellcolor[HTML]{FFD966}                        & \cellcolor[HTML]{FFD966}                        & \cellcolor[HTML]{FFD966}                        & \cellcolor[HTML]{FFD966}                        & \cellcolor[HTML]{FFD966}                        & \cellcolor[HTML]{FFD966}                        & \cellcolor[HTML]{FFD966}                        & \cellcolor[HTML]{FFD966}                        & \cellcolor[HTML]{FFD966}                        & \cellcolor[HTML]{FFD966}                        & \cellcolor[HTML]{FFD966}                        & \cellcolor[HTML]{FFD966}                        & \cellcolor[HTML]{FFD966}                        & \cellcolor[HTML]{FFD966}                        & \cellcolor[HTML]{FFD966}                        & \cellcolor[HTML]{FFD966}                        & \cellcolor[HTML]{FFD966}                        & \cellcolor[HTML]{FFD966}                        & \cellcolor[HTML]{FFD966}                        & \cellcolor[HTML]{FFD966} & \cellcolor[HTML]{FFD966} & \cellcolor[HTML]{FFD966} & \cellcolor[HTML]{FFD966} & \cellcolor[HTML]{FFD966} & \cellcolor[HTML]{FFD966} & \cellcolor[HTML]{FFD966} & \cellcolor[HTML]{FFD966} \\ \hhline{|~|-|-|------------------|-----|-------|----|------------------|} 
                                            &                                                   & G & \cellcolor[HTML]{F4B083}                        & \cellcolor[HTML]{F4B083}                        & \cellcolor[HTML]{F4B083}                        & \cellcolor[HTML]{F4B083}                        & \cellcolor[HTML]{F4B083}                        & \cellcolor[HTML]{F4B083}                        & \cellcolor[HTML]{F4B083}                        & \cellcolor[HTML]{F4B083}                        & \cellcolor[HTML]{F4B083}                        & \cellcolor[HTML]{F4B083}                        & \cellcolor[HTML]{F4B083}                        & \cellcolor[HTML]{F4B083}                        & \cellcolor[HTML]{F4B083}                        & \cellcolor[HTML]{F4B083}                        & \cellcolor[HTML]{F4B083}                        & \cellcolor[HTML]{F4B083}                        & \cellcolor[HTML]{F4B083}                        & \cellcolor[HTML]{F4B083}                        & \cellcolor[HTML]{F4B083}                        & \cellcolor[HTML]{F4B083}                        & \cellcolor[HTML]{F4B083}                        & \cellcolor[HTML]{F4B083}                        & \cellcolor[HTML]{F4B083}                        & \cellcolor[HTML]{F4B083}                        & \cellcolor[HTML]{F4B083}                        & \cellcolor[HTML]{F4B083}                        & \cellcolor[HTML]{F4B083}                        & \cellcolor[HTML]{F4B083}                        & \cellcolor[HTML]{F4B083}                        & \cellcolor[HTML]{F4B083}                        & \cellcolor[HTML]{F4B083}                        & \cellcolor[HTML]{F4B083}                        & \cellcolor[HTML]{F4B083}                        & \cellcolor[HTML]{F4B083}                        & \cellcolor[HTML]{F4B083}                        & \cellcolor[HTML]{F4B083}                        & \cellcolor[HTML]{F4B083}                        & \cellcolor[HTML]{F4B083}                        & \cellcolor[HTML]{F4B083}                        & \cellcolor[HTML]{F4B083}                        & \cellcolor[HTML]{F4B083}                        & \cellcolor[HTML]{F4B083}                        & \cellcolor[HTML]{F4B083}                        & \cellcolor[HTML]{F4B083}                        & \cellcolor[HTML]{F4B083} & \cellcolor[HTML]{F4B083} & \cellcolor[HTML]{F4B083} & \cellcolor[HTML]{F4B083} & \cellcolor[HTML]{F4B083} & \cellcolor[HTML]{F4B083} & \cellcolor[HTML]{F4B083} & \cellcolor[HTML]{F4B083} \\ \hhline{|~|~|-|~~~~~~~~~~~~~~~~~~|~~~~~|~~~~~~~|~~~~|~~~~~~~~~~~~~~~~~~|}
                                            &                                                   & S & \cellcolor[HTML]{F4B083}                        & \cellcolor[HTML]{F4B083}                        & \cellcolor[HTML]{F4B083}                        & \cellcolor[HTML]{F4B083}                        & \cellcolor[HTML]{F4B083}                        & \cellcolor[HTML]{F4B083}                        & \cellcolor[HTML]{F4B083}                        & \cellcolor[HTML]{F4B083}                        & \cellcolor[HTML]{F4B083}                        & \cellcolor[HTML]{F4B083}                        & \cellcolor[HTML]{F4B083}                        & \cellcolor[HTML]{F4B083}                        & \cellcolor[HTML]{F4B083}                        & \cellcolor[HTML]{F4B083}                        & \cellcolor[HTML]{F4B083}                        & \cellcolor[HTML]{F4B083}                        & \cellcolor[HTML]{F4B083}                        & \cellcolor[HTML]{F4B083}                        & \cellcolor[HTML]{F4B083}                        & \cellcolor[HTML]{F4B083}                        & \cellcolor[HTML]{F4B083}                        & \cellcolor[HTML]{F4B083}                        & \cellcolor[HTML]{F4B083}                        & \cellcolor[HTML]{F4B083}                        & \cellcolor[HTML]{F4B083}                        & \cellcolor[HTML]{F4B083}                        & \cellcolor[HTML]{F4B083}                        & \cellcolor[HTML]{F4B083}                        & \cellcolor[HTML]{F4B083}                        & \cellcolor[HTML]{F4B083}                        & \cellcolor[HTML]{F4B083}                        & \cellcolor[HTML]{F4B083}                        & \cellcolor[HTML]{F4B083}                        & \cellcolor[HTML]{F4B083}                        & \cellcolor[HTML]{F4B083}                        & \cellcolor[HTML]{F4B083}                        & \cellcolor[HTML]{F4B083}                        & \cellcolor[HTML]{F4B083}                        & \cellcolor[HTML]{F4B083}                        & \cellcolor[HTML]{F4B083}                        & \cellcolor[HTML]{F4B083}                        & \cellcolor[HTML]{F4B083}                        & \cellcolor[HTML]{F4B083}                        & \cellcolor[HTML]{F4B083}                        & \cellcolor[HTML]{F4B083} & \cellcolor[HTML]{F4B083} & \cellcolor[HTML]{F4B083} & \cellcolor[HTML]{F4B083} & \cellcolor[HTML]{F4B083} & \cellcolor[HTML]{F4B083} & \cellcolor[HTML]{F4B083} & \cellcolor[HTML]{F4B083} \\ \hhline{|~|~|-|~~~~~~~~~~~~~~~~~~|~~~~~|~~~~~~~|~~~~|~~~~~~~~~~~~~~~~~~|}
                                            & \multirow{-3}{*}{\textbf{I}}                      & P & \cellcolor[HTML]{FFD966}                        & \cellcolor[HTML]{F4B083}                        & \cellcolor[HTML]{F4B083}                        & \cellcolor[HTML]{F4B083}                        & \cellcolor[HTML]{F4B083}                        & \cellcolor[HTML]{F4B083}                        & \cellcolor[HTML]{F4B083}                        & \cellcolor[HTML]{F4B083}                        & \cellcolor[HTML]{F4B083}                        & \cellcolor[HTML]{F4B083}                        & \cellcolor[HTML]{F4B083}                        & \cellcolor[HTML]{F4B083}                        & \cellcolor[HTML]{F4B083}                        & \cellcolor[HTML]{F4B083}                        & \cellcolor[HTML]{F4B083}                        & \cellcolor[HTML]{F4B083}                        & \cellcolor[HTML]{F4B083}                        & \cellcolor[HTML]{F4B083}                        & \cellcolor[HTML]{F4B083}                        & \cellcolor[HTML]{F4B083}                        & \cellcolor[HTML]{F4B083}                        & \cellcolor[HTML]{F4B083}                        & \cellcolor[HTML]{F4B083}                        & \cellcolor[HTML]{F4B083}                        & \cellcolor[HTML]{F4B083}                        & \cellcolor[HTML]{F4B083}                        & \cellcolor[HTML]{F4B083}                        & \cellcolor[HTML]{F4B083}                        & \cellcolor[HTML]{F4B083}                        & \cellcolor[HTML]{F4B083}                        & \cellcolor[HTML]{F4B083}                        & \cellcolor[HTML]{F4B083}                        & \cellcolor[HTML]{F4B083}                        & \cellcolor[HTML]{F4B083}                        & \cellcolor[HTML]{F4B083}                        & \cellcolor[HTML]{F4B083}                        & \cellcolor[HTML]{F4B083}                        & \cellcolor[HTML]{F4B083}                        & \cellcolor[HTML]{F4B083}                        & \cellcolor[HTML]{F4B083}                        & \cellcolor[HTML]{F4B083}                        & \cellcolor[HTML]{F4B083}                        & \cellcolor[HTML]{F4B083}                        & \cellcolor[HTML]{F4B083}                        & \cellcolor[HTML]{F4B083} & \cellcolor[HTML]{F4B083} & \cellcolor[HTML]{F4B083} & \cellcolor[HTML]{F4B083} & \cellcolor[HTML]{F4B083} & \cellcolor[HTML]{F4B083} & \cellcolor[HTML]{F4B083} & \cellcolor[HTML]{F4B083} \\ \hhline{|~|-|-|------------------|-----|-------|----|------------------|} 
                                            &                                                   & G & \cellcolor[HTML]{F4B083}                        & \cellcolor[HTML]{F4B083}                        & \cellcolor[HTML]{F4B083}                        & \cellcolor[HTML]{F4B083}                        & \cellcolor[HTML]{F4B083}                        & \cellcolor[HTML]{F4B083}                        & \cellcolor[HTML]{F4B083}                        & \cellcolor[HTML]{F4B083}                        & \cellcolor[HTML]{F4B083}                        & \cellcolor[HTML]{F4B083}                        & \cellcolor[HTML]{F4B083}                        & \cellcolor[HTML]{F4B083}                        & \cellcolor[HTML]{F4B083}                        & \cellcolor[HTML]{F4B083}                        & \cellcolor[HTML]{F4B083}                        & \cellcolor[HTML]{F4B083}                        & \cellcolor[HTML]{F4B083}                        & \cellcolor[HTML]{F4B083}                        & \cellcolor[HTML]{F4B083}                        & \cellcolor[HTML]{F4B083}                        & \cellcolor[HTML]{F4B083}                        & \cellcolor[HTML]{F4B083}                        & \cellcolor[HTML]{F4B083}                        & \cellcolor[HTML]{F4B083}                        & \cellcolor[HTML]{F4B083}                        & \cellcolor[HTML]{F4B083}                        & \cellcolor[HTML]{F4B083}                        & \cellcolor[HTML]{F4B083}                        & \cellcolor[HTML]{F4B083}                        & \cellcolor[HTML]{F4B083}                        & \cellcolor[HTML]{F4B083}                        & \cellcolor[HTML]{F4B083}                        & \cellcolor[HTML]{F4B083}                        & \cellcolor[HTML]{F4B083}                        & \cellcolor[HTML]{F4B083}                        & \cellcolor[HTML]{F4B083}                        & \cellcolor[HTML]{F4B083}                        & \cellcolor[HTML]{F4B083}                        & \cellcolor[HTML]{F4B083}                        & \cellcolor[HTML]{F4B083}                        & \cellcolor[HTML]{F4B083}                        & \cellcolor[HTML]{F4B083}                        & \cellcolor[HTML]{F4B083}                        & \cellcolor[HTML]{F4B083}                        & \cellcolor[HTML]{F4B083} & \cellcolor[HTML]{F4B083} & \cellcolor[HTML]{F4B083} & \cellcolor[HTML]{F4B083} & \cellcolor[HTML]{F4B083} & \cellcolor[HTML]{F4B083} & \cellcolor[HTML]{F4B083} & \cellcolor[HTML]{F4B083} \\ \hhline{|~|~|-|~~~~~~~~~~~~~~~~~~|~~~~~|~~~~~~~|~~~~|~~~~~~~~~~~~~~~~~~|}
                                            &                                                   & S & \cellcolor[HTML]{A8D08D}                        & \cellcolor[HTML]{A8D08D}                        & \cellcolor[HTML]{A8D08D}                        & \cellcolor[HTML]{A8D08D}                        & \cellcolor[HTML]{A8D08D}                        & \cellcolor[HTML]{A8D08D}                        & \cellcolor[HTML]{A8D08D}                        & \cellcolor[HTML]{A8D08D}                        & \cellcolor[HTML]{A8D08D}                        & \cellcolor[HTML]{A8D08D}                        & \cellcolor[HTML]{A8D08D}                        & \cellcolor[HTML]{A8D08D}                        & \cellcolor[HTML]{A8D08D}                        & \cellcolor[HTML]{A8D08D}                        & \cellcolor[HTML]{A8D08D}                        & \cellcolor[HTML]{A8D08D}                        & \cellcolor[HTML]{A8D08D}                        & \cellcolor[HTML]{A8D08D}                        & \cellcolor[HTML]{A8D08D}                        & \cellcolor[HTML]{A8D08D}                        & \cellcolor[HTML]{A8D08D}                        & \cellcolor[HTML]{A8D08D}                        & \cellcolor[HTML]{A8D08D}                        & \cellcolor[HTML]{A8D08D}                        & \cellcolor[HTML]{A8D08D}                        & \cellcolor[HTML]{A8D08D}                        & \cellcolor[HTML]{A8D08D}                        & \cellcolor[HTML]{A8D08D}                        & \cellcolor[HTML]{A8D08D}                        & \cellcolor[HTML]{A8D08D}                        & \cellcolor[HTML]{A8D08D}                        & \cellcolor[HTML]{A8D08D}                        & \cellcolor[HTML]{A8D08D}                        & \cellcolor[HTML]{A8D08D}                        & \cellcolor[HTML]{A8D08D}                        & \cellcolor[HTML]{A8D08D}                        & \cellcolor[HTML]{A8D08D}                        & \cellcolor[HTML]{A8D08D}                        & \cellcolor[HTML]{A8D08D}                        & \cellcolor[HTML]{A8D08D}                        & \cellcolor[HTML]{A8D08D}                        & \cellcolor[HTML]{A8D08D}                        & \cellcolor[HTML]{A8D08D}                        & \cellcolor[HTML]{A8D08D}                        & \cellcolor[HTML]{A8D08D} & \cellcolor[HTML]{A8D08D} & \cellcolor[HTML]{A8D08D} & \cellcolor[HTML]{A8D08D} & \cellcolor[HTML]{A8D08D} & \cellcolor[HTML]{A8D08D} & \cellcolor[HTML]{A8D08D} & \cellcolor[HTML]{A8D08D} \\ \hhline{|~|~|-|~~~~~~~~~~~~~~~~~~|~~~~~|~~~~~~~|~~~~|~~~~~~~~~~~~~~~~~~|}
\multirow{-15}{*}{\rotatebox[origin=c]{90}{\textbf{CONTROL GROUP}}}   & \multirow{-3}{*}{\textbf{J}}                      & P & \cellcolor[HTML]{FFD966}                        & \cellcolor[HTML]{F4B083}                        & \cellcolor[HTML]{F4B083}                        & \cellcolor[HTML]{F4B083}                        & \cellcolor[HTML]{F4B083}                        & \cellcolor[HTML]{F4B083}                        & \cellcolor[HTML]{F4B083}                        & \cellcolor[HTML]{F4B083}                        & \cellcolor[HTML]{F4B083}                        & \cellcolor[HTML]{F4B083}                        & \cellcolor[HTML]{F4B083}                        & \cellcolor[HTML]{F4B083}                        & \cellcolor[HTML]{F4B083}                        & \cellcolor[HTML]{F4B083}                        & \cellcolor[HTML]{F4B083}                        & \cellcolor[HTML]{F4B083}                        & \cellcolor[HTML]{F4B083}                        & \cellcolor[HTML]{F4B083}                        & \cellcolor[HTML]{F4B083}                        & \cellcolor[HTML]{F4B083}                        & \cellcolor[HTML]{F4B083}                        & \cellcolor[HTML]{F4B083}                        & \cellcolor[HTML]{F4B083}                        & \cellcolor[HTML]{F4B083}                        & \cellcolor[HTML]{F4B083}                        & \cellcolor[HTML]{F4B083}                        & \cellcolor[HTML]{F4B083}                        & \cellcolor[HTML]{F4B083}                        & \cellcolor[HTML]{F4B083}                        & \cellcolor[HTML]{F4B083}                        & \cellcolor[HTML]{F4B083}                        & \cellcolor[HTML]{F4B083}                        & \cellcolor[HTML]{F4B083}                        & \cellcolor[HTML]{F4B083}                        & \cellcolor[HTML]{F4B083}                        & \cellcolor[HTML]{F4B083}                        & \cellcolor[HTML]{F4B083}                        & \cellcolor[HTML]{F4B083}                        & \cellcolor[HTML]{F4B083}                        & \cellcolor[HTML]{F4B083}                        & \cellcolor[HTML]{F4B083}                        & \cellcolor[HTML]{F4B083}                        & \cellcolor[HTML]{F4B083}                        & \cellcolor[HTML]{F4B083}                        & \cellcolor[HTML]{F4B083} & \cellcolor[HTML]{F4B083} & \cellcolor[HTML]{F4B083} & \cellcolor[HTML]{F4B083} & \cellcolor[HTML]{F4B083} & \cellcolor[HTML]{F4B083} & \cellcolor[HTML]{F4B083} & \cellcolor[HTML]{F4B083} \\ \hline
\end{tabu}
\end{table*}

\textbf{GPS.} In our treatment group, the behaviour change was dramatic. Across the five individuals, GPS usage decreased by an average of 40\%. Our small sample impeded significance (\textit{p} = 0.157), but the `very large' effect size (\textit{d} = 1.461) was promising \cite{Sawilowsky2009}. Based on a desired power of 0.8, a sample of at least 14 would be required. While Participants \textit{B} and \textit{D} used the service in the pretest phase, they ceased usage during the gameplay session. Furthermore, they did not re-enable GPS throughout the remainder of the study. This implies that users successfully learned protection. 

This comes in contrast to the control group, where every participant allowed the service. Usage did not adjust even slightly between their pretest and posttest phases (\textit{p} = 1.0). Indeed, GPS was not disabled once over the 52 days. This suggests that, without training, users will not protect themselves. It also implies that behaviour was not biased by our questionnaires.

\textbf{Screen locks.} For the treatment group, screen lock usage increased by 42.7\%. This change was significant (\textit{Z} = -2.023, \textit{p} = 0.043, \textit{d} = 0.733), with the medium effect size suggesting the game was persuasive. Participants \textit{B} and \textit{E} did not use a password during the first 18 days. However, within 10 minutes of playing the game, both enabled the feature.

In the control group, Participant \textit{J} continued to use a password from the pretest stage. None of the other individuals used the feature even once. As such, behaviour barely changed as the study progressed (\textit{p} = 0.317). Again, this demonstrates that protection will be rarely used without encouragement.

\textbf{Permissions.} Interestingly, permission acceptance did not differ greatly for either group. In our treatment group (\textit{p} = 0.5), the acceptance rate was stable for \textit{A}, \textit{D} and \textit{E}. \textit{B} might have responded to the game, decreasing their percentage by 22\%. However, \textit{C} continued their exploration, increasing their rate by 21.5\%. Users often spoke of balancing privacy against functionality, and these views are explored in the next section. 

In the control group (\textit{p} = 0.068), behaviour remained stable for all individuals. This suggests that their game did not influence privacy protection. Across the 52 days, no control participants revoked a single permission. Their \textit{permission scores} only differed based on the apps they installed. This further implies that protective behaviour is rare on smartwatches. 

\textbf{RQ3.} Table \ref{fig:time} illustrates protective behaviour throughout the study. As shown, control-group actions are static before, during and after gameplay. Indeed, actions appeared finalised from Day 2 of the study. Throughout the other 50 days, the shading continues to be orange. Therefore, it appears that the generic game had no influence on behaviour. This ensures that it served as an appropriate control to the privacy app.

For the treatment group, protection was rare in the pretest period. Although some participants used the features, their usage was inconsistent. However, once the gameplay phase begins, the chart becomes predominantly green and yellow. This shading remains throughout the rest of the study. As the privacy game lost salience, behaviour did not appear to relapse. This suggests that this app was successful in encouraging protection. As concerns decreased in the treatment group, opinions and behaviour appeared to realign.

\subsection{Interview Findings}

After the posttest questionnaires were completed, we concluded the study with interviews. The questions can be found in Table \ref{tbl:interview} of the appendices. These semi-structured discussions served three purposes. Firstly, they allowed us to gauge general opinions of the study. Secondly, we could compare the privacy knowledge of our two groups. Finally, we explored the behavioural rationale of each participant. As an overview, the responses of our groups can be found below in Table \ref{tbl:results}. This illustrates how the posttest capabilities of participants appeared to differ after the gameplay phase. The details of the responses are highlighted within the following paragraphs.

\begin{table}[H]
\centering
\caption{Participant interview results}
\label{tbl:results}
\bgroup
\def\arraystretch{1.3}
\begin{tabular}{|l|l|c|c|}
\hline
\textbf{Q}  & \multicolumn{1}{c|}{\textbf{Characteristic}} & \textbf{Treatment (\%)}              & \textbf{Control (\%)}                \\ \hline
\textbf{10} & \textbf{Knew protective feature}             & \cellcolor[HTML]{DAF0DD}\textbf{100} & \cellcolor[HTML]{FFECCD}\textbf{40}  \\ \cline{3-4} 
            & Disable GPS                                  & \cellcolor[HTML]{FFFECD}60           & \cellcolor[HTML]{FCE4D6}20           \\ \cline{3-4} 
            & Lock screen                                  & \cellcolor[HTML]{FFECCD}40           & \cellcolor[HTML]{FCD6D6}0            \\ \cline{3-4} 
            & Restrict permissions                         & \cellcolor[HTML]{FFFECD}60           & \cellcolor[HTML]{FCE4D6}20           \\ \hline
            & \textbf{Would defend privacy}                & \cellcolor[HTML]{DAF0DD}\textbf{100} & \cellcolor[HTML]{FFFECD}\textbf{60}  \\ \cline{3-4} 
\textbf{11} & Against location tracking                    & \cellcolor[HTML]{DAF0DD}100          & \cellcolor[HTML]{FCD6D6}0            \\ \cline{3-4} 
\textbf{12} & Against app data collection                  & \cellcolor[HTML]{DAF0DD}100          & \cellcolor[HTML]{FCD6D6}0            \\ \cline{3-4} 
\textbf{13} & Against unauthorised access                  & \cellcolor[HTML]{DAF0DD}100          & \cellcolor[HTML]{FFFECD}60           \\ \hline
            & \textbf{Demonstrated ability}                & \cellcolor[HTML]{DAF0DD}\textbf{100} & \cellcolor[HTML]{DAF0DD}\textbf{100} \\ \cline{3-4} 
\textbf{14} & Disabling GPS                                & \cellcolor[HTML]{DAF0DD}100          & \cellcolor[HTML]{FFFECD}60           \\ \cline{3-4} 
\textbf{15} & Locking screen                               & \cellcolor[HTML]{ECEFDB}80           & \cellcolor[HTML]{ECEFDB}80           \\ \cline{3-4} 
\textbf{16} & Restricting permissions                      & \cellcolor[HTML]{ECEFDB}80           & \cellcolor[HTML]{ECEFDB}80           \\ \hline
\textbf{}   & \textbf{Demonstrated confidence}             & \cellcolor[HTML]{DAF0DD}\textbf{100} & \cellcolor[HTML]{FFECCD}\textbf{40}  \\ \cline{3-4} 
\textbf{14} & Disabling GPS                                & \cellcolor[HTML]{DAF0DD}100          & \cellcolor[HTML]{FCE4D6}20           \\ \cline{3-4} 
\textbf{15} & Restricting permissions                      & \cellcolor[HTML]{DAF0DD}100          & \cellcolor[HTML]{FFECCD}40           \\ \cline{3-4} 
\textbf{16} & Locking screen                               & \cellcolor[HTML]{ECEFDB}80           & \cellcolor[HTML]{FCD6D6}0            \\ \hline
\end{tabular}
\egroup
\end{table}

\textbf{General.} We asked users whether they felt influenced by the background monitoring app. None of our 10 participants believed it had any effect. While this does not ensure external validity, it increases the reliability of our findings.

\begin{quoting}
``\textit{I just used it as I would normally}'' (\#A, Treatment).
\end{quoting}

With the study requiring long-term interaction, we were interested in why our users chose to participate. All 10 were curious to trial a smartwatch, with two also appreciating research. Only two mentioned the voucher compensation, suggesting participation was primarily driven by genuine interest. Since our demographics were not dissimilar to the user population \cite{NPDConnectedIntelligence2014,Desarnauts2016}, we should have external validity.

\begin{quoting}
``\textit{I was about to buy a new one [smartwatch] so that was the perfect moment}'' (\#J, Control).
\end{quoting}

Before introducing the topic of privacy, we asked participants if they learned anything as the study progressed. This assessed gameplay retention, as the apps had not been used for 18 days. Privacy was highlighted by 60\% of the treatment group. Since they also praised the game, it might have been educational. The control group were similarly influenced, with 60\% mentioning their app. However, to truly examine whether users are informed, we must test their knowledge.

\begin{quoting}
``\textit{I think there was a couple of privacy settings that, through the game, I picked up}'' (\#B, Treatment).
\end{quoting}

\textbf{Privacy awareness.} To compare degrees of privacy awareness, we asked users how they believed their data could be accessed. All of our treatment group provided accurate descriptions (100\%). They also highlighted the risk of user accounts (12.5\%) and fraudulent apps (5.0\%). Since their game outlined privacy threats, they might have learned of their vulnerability. In contrast, only 40\% of controls knew app practices, with the others blaming irrelevant technologies.

\begin{quoting}
``\textit{Through some app that you allow them to track your location}'', (\#E, Treatment).
\end{quoting}

We then asked users how they could protect their privacy. As shown in Table \ref{tbl:results}, all treatment participants knew a beneficial action (100\%). 40\% named screen locks, 60\% cited permissions and 60\% would disable GPS. Even if they choose not to act, they should be able to make informed decisions. In the control group, only 40\% named a single setting. Many justifications highlighted that they were unsure (13.3\%) or unconcerned (13.3\%). If individuals lack awareness, their data might be placed at risk.

\begin{quoting}
``\textit{I'd probably start by going through the list of apps and seeing what permissions were useless}'', (\#A, Treatment).
\end{quoting}

To further assess knowledge, we asked participants how they could defend against the scenarios. These comprised of: location tracking, unauthorised access and app data collection. If individuals know defences, they can act in response to their concerns. In the first incident, all the treatment group knew to disable their GPS (100\%). However, none of the control participants could list a technique (0\%). When considering unauthorised access, the former group also performed well. All the users mentioned screen locks, whether PIN (60\%) or pattern (40\%). This was compared with 60\% of the controls, with several highlighting they felt unsure. Finally, we considered defences against data collection. The treatment group outlined permissions (100\%) and app deletion (60\%). No control participants knew of permissions, even after 52 days of interaction (0\%). When comparing the groups, it appears as if the privacy game was educational. Unfortunately, untrained users seem not to seek out protection.

\begin{quoting}
``\textit{...I'm not using Google Maps right now, so I don't need to have the location enabled for it}'' (\#D - TMT).
\end{quoting}

\textbf{Ability.} While these responses gave us confidence, we wished to test knowledge empirically. Individuals might know of settings but be unable to use them. Therefore, we asked users to demonstrate the three protective features. By talking aloud, we could ascertain both their route and their certainty.

Users were given a watch and asked to disable the GPS. The treatment group found this simple, with all five navigating directly (100\%). In contrast, only 60\% of control participants could find the settings. Another 40\% claimed to have never checked the feature, indicating their lack of exploration. Individuals were then asked to adjust their permissions. These settings were better-understood, with 8/10 navigating straight to the menu. However, while the treatment group were all certain (100\%), 60\% of the others were learning en route. 

\begin{quoting}
``\textit{Disable GPS you said? Go down to Connectivity, Location, off. Done}'', (\#A, Treatment).
\end{quoting}

When requesting password usage, individuals had greater difficulty. Although 7/10 followed a direct path, no control participants expressed certainty (0\%). This appeared due to the difficulty in categorising \textit{Screen Lock} in a particular menu. After 52 days of interaction, it is concerning that privacy settings cause such confusion. This further demonstrates the importance of educational tools.

\begin{quoting}
``\textit{Another thing I haven't done}'' (\#I, Control).
\end{quoting}

\subsection{What factors influence smartwatch behaviour?}

\textbf{RQ4.} Finally, we consider the responses to our PMT questions. Based on the frequency of themes, we outline the factors that appear most influential. Through exploring user rationale, we address our final research question.

\textbf{PMT factors.} In terms of factors, participants generally possessed good self-efficacy. Most also believed configuration was easy, even if some knowledge was required. And although many in the control group doubted watch protection, settings did not seem to be the issue. 

The most influential factors appeared to be the threat components. Firstly, most users had a balanced view of severity. If data access was consented and rewarded, many were satisfied. This helps explain why protection was so often ignored. Secondly, the control group failed to perceive risk. Unlike treatment participants, they had not learned the value of their data. Finally, and most influentially, users received rewards from smartwatch apps. Since settings can impede functionality, permissions were often blindly accepted. It was only after gameplay that participants reflected on data access.

\textbf{Rationale.} Based on interview responses, three issues primarily influenced decisions: sensitivity, salience and convenience. When individuals knew that their data was valuable, they considered protection. However, since watch details were often deemed innocuous, settings were not explored. Similarly, when privacy was not visible, participants often forgot the concept. If they felt at risk or noticed consequences, protection regained its relevance. Most crucially, users tended to weigh convenience against privacy. Smartwatches are obtained to provide functionality, and settings can restrict these benefits. Therefore, even informed users would trade some data, while actively protecting other details.

\textbf{Persuasion.} Through our final questions, we asked participants what would encourage protection. Users named a range of scenarios, from negative media reports to apps being hacked. A common theme was if the participant acquired a high-profile job. This would increase the sensitivity of watch data, and hence encourage greater protection. Individuals also claimed they would act if abroad, especially if that country was dangerous. However, one participant did go overseas, and reported not increasing their protection. Privacy settings were rarely mentioned as an impediment. As before, this suggests that threat components have greater influence on smartwatches.

\textbf{Further approaches.} The largest issues appear to be perceived severity, perceived vulnerability and perceived rewards. Severity is challenging to magnify, since many participants had a nuanced view. Individuals did not oppose all sharing, though perceptions might change after recent privacy controversies \cite{Glenday2018}. To highlight the risk from their data, we could demonstrate inference techniques through online videos.

Our privacy game contextualised challenges around real-life behaviour. This sought to make the issue salient to each individual. After treatment participants learned of their vulnerability, many chose to adjust their settings. Since risk appears influential, future tools could analyse user permissions. Based on the restrictions applied to each app, a risk exposure could be calculated. By allowing individuals to compare their scores, protection might be incentivised.

Rewards are challenging to counter, since apps do provide convenient features. However, this does not mean that data has to be sacrificed. Mocking frameworks have been successful in faking smartphone metrics \cite{Beresford2011}. Since our watches run on an Android environment, similar tools might be possible. When an app then requires a location, a coarse position could be given. In this manner, functionality could be received while protecting data.

\section{Conclusions and Further Work}
\label{sec:six}

\textbf{Summary.} We outline the development of the first privacy-themed smartwatch game. It was designed through Learning Science principles and evaluated through a 52-day longitudinal study. Our treatment group, who played the game, began taking greater action to protect their privacy. Indeed, their usage of screen locks significantly increased after gameplay. The control group, who used a generic version, continued to do little. Indeed, 80\% of these users failed to adjust a single setting. Since treatment concerns became more nuanced after gameplay, opinions appeared to realign with behaviour.

By dissecting interviews through Protection Motivation Theory, we explored smartwatch privacy rationale. Participants appeared most influenced by three factors: sensitivity, salience and convenience. A person will not invest effort unless their data is deemed valuable. Even if they do desire protection, privacy can be easily overlooked. Finally, informed users might sacrifice data for convenience. However, they can only make a considered choice if they understand the risks. Since smartwatch games appear to encourage protection, they should be considered as a complement to awareness campaigns.

\textbf{Implications.} Our findings are in line with existing research. As highlighted above, even informed users might trade their data for functionality. This supports the concept of Privacy Calculus, where the benefits and risks of disclosure are compared \cite{Culnan1999}. However, until individuals gain an understanding of the topic, they cannot judge the risk fairly \cite{Slovic1987}. Indeed, as highlighted by Acquisti et al. \citeyear{Acquisti2015}, the Privacy Paradox ``\textit{is also affected by misperceptions of those costs and benefits, as well as social norms, emotions, and heuristics}''. When people lack knowledge of a matter, they tend to overestimate the advantages \cite{Gomez2018}. Therefore, since baseline privacy knowledge tends to be low \cite{Bashir2015}, we must support users to make informed protective decisions. We believe this has been achieved through the use of educational games. As suggested by Hallam and Zanella \citeyear{Hallam2017}, privacy issues became more pertinent after we increased their salience.

\textbf{Permanence.} Although interventions might adjust behaviour, they can lose efficacy once their salience decreases. As participants forget about our educational game, they might decrease their protective behaviour. We sought to influence the availability heuristic as a means of increasing risk perception \cite{Tversky1973}. Therefore, we recognise that as salience reduces, so does the perceived likelihood of threats. However, our game also aimed to enhance individuals' self-efficacy. Even if participants lack the immediate desire to guard their data, it is valuable that they know how. Our posttest interviews showed that even the treatment participants who avoided protection (e.g., Participant C) could demonstrate the usage of privacy settings. Hence, although salience might decrease over the longer-term, protective knowledge should be retained. This was suggested in our posttest results, where behaviour did not revert even weeks after gameplay.

\textbf{Limitations.} We are transparent in the fact that our study possesses several limitations. Firstly, we only evaluated a sample size of 10 participants. As a result, we drew no conclusions over whether privacy concerns vary by culture. We were constrained, since new watches required monitoring over a consistent period. While we would have preferred a larger group, we supported our quantitative findings with a rich qualitative analysis. Secondly, our longitudinal study only spanned a period of two months. In this case, we were limited by the term lengths of our university. However, the duration was in excess of many two-stage studies, which impose a gap of one week \cite{Albayram2017,Wiedenbeck2005,DeWitt2006,Kumaraguru2009}.
To test retention further, we plan to monitor new participants over a longer period. Thirdly, our gameplay questionnaire suggested that only 60\% of users enjoyed the games. Therefore, even if the privacy app did prove beneficial, further enhancements might be required to support intrinsic motivation. Finally, by targeting Android devices, we neglected the Apple Watch environment. This is a distinct ecosystem, albeit one which may be less amenable to analysis \cite{Tracy2012}. In future research, we seek to compare behaviour by developing Apple games.

\textbf{Further work.} We finally discuss opportunities for further work. It would be interesting to explore smartwatch purchases. Wearables may present risks, but it is unclear whether this fact is ever considered. By comparing Wear OS users to other populations, we could analyse how expectations vary. Smartwatch games appear to be effective in encouraging protection. However, other connected devices, such as Smart TVs, also present privacy issues \cite{Ghiglieri2017}. Our design principles are transferable, and TV games might highlight the risk.

\bibliographystyle{apacite}
\bibliography{bib}

\newpage

\appendix

\vspace{-2em}

\begin{table}[h!]
\caption{Concern Questionnaire: Privacy Questions in Bold}
\footnotesize
\setlength{\tabcolsep}{.25em}
\begin{tabular}{m{0.5cm} p{7.3cm}} 
\toprule
\# & Indicate your agreement or disagreement with the statements. \\
& Strongly Agree, Agree, Neutral, Disagree, Strongly Disagree.\\
\midrule
1 & ``I find the smartwatch useful.''
 \\[1ex]
2 & ``I use a wide range of features on the smartwatch.''
 \\[1ex]
3 & ``I would experience inconvenience if I didn't use the smartwatch.''
 \\[1ex]
4 & ``It is possible for smartwatch apps to simplify common tasks.''
 \\[1ex] 
5 & \textbf{``It is possible for smartwatch apps to access personal data.''}
 \\[1ex] 
6 & ``It is possible for smartwatch apps to drain the battery.''
 \\[1ex]
7 & ``I have a strong understanding of smartwatch notification features.''
 \\[1ex]
8 & \textbf{``I have a strong understanding of smartwatch privacy features.''}
 \\[1ex]
9 & It is important you remain attentive. Indicate that you are by marking X in the Strongly Disagree box.
 \\[1ex]
10 & ``I have a strong understanding of smartwatch display features.''
 \\[1ex]
11 & \textbf{``There is a realistic chance of smartwatches being lost or stolen.''}
 \\[1ex] 
12 & ``If I didn't configure my settings, my apps might drain my battery.''
 \\[1ex]
13 & \textbf{``If I didn't configure my settings, my apps might place my data at risk.''}
 \\[1ex]
14 & ``If I didn't configure my settings, my apps might slow down my watch.''
 \\\midrule
 \# & Indicate your responses from Indifferent to Very Concerned. \\
 & Also provide your qualitative rationale.\\
 \midrule
15 & How would you feel if Google (the developer of Android) changed your smartwatch's default font size?
 \\[1ex]
16 & \textbf{How would you feel if app companies could track your precise current location?}
 \\[1ex]
17 & Imagine a software update changed your smartwatch's font size. How would you feel if the text was made much smaller than it was before?
 \\[1ex]
18 & \textbf{How would you feel if app companies could read your personal data from your smartwatch?}
 \\[1ex] 
19 & \textbf{Imagine your smartwatch was lost or stolen. How would you feel if a random stranger could read your data?}
 \\[1ex] 
20 & How would you feel if Google (the developer of Android) changed your smartwatch's default alarm volume?
 \\[1ex]
21 & \textbf{Imagine your smartwatch was lost or stolen. How would you feel if a random stranger could use your apps as you?}
 \\[1ex]
22 & Imagine a software update changed your smartwatch's alarm volume. How would you feel if the alarm volume was set much quieter than it was before?
 \\[1ex]
23 & \textbf{How would you feel if app companies could share your personal data with other companies?}
 \\[1ex]
24 & How would you feel if Google  (the developer of Android) changed your smartwatch's default screen brightness?
 \\[1ex]
25 & \textbf{How would you feel if app companies could share your precise movements with other companies?}
 \\[1ex] 
26 & Imagine a software update changed your smartwatch's screen brightness. How would you feel if the brightness was set much lower than it was before?
 \\
\bottomrule
\end{tabular}
\centering
\label{tbl:concerns}
\end{table}

\begin{table}[H]
\caption[]{Posttest Interview Questions:\\All Questions Solicit Open-Ended Responses}
\footnotesize
\setlength{\tabcolsep}{.25em}
\begin{tabular}{m{0.5cm} p{7.3cm}} 
\toprule
\# & Introductory Questions \\ 
\midrule
1 & What was your experience in wearing the smartwatch?
 \\[1ex]
2 & Why did you choose to participate in the study?
 \\[1ex]
3 & Do you feel the background StudyService app affected your behaviour? Why?
 \\[1ex]
4 & Would you purchase your own smartwatch? Why?
 \\[1ex] 
5 & Do you feel you learned anything new as the study progressed? If so, what?
  \\\midrule
 \# & Privacy Awareness and Knowledge Questions \\
 \midrule
6 & How likely do you believe the chance of companies accessing your watch's data? Why?
 \\[1ex]
7 & How likely do you believe the chance of someone's smartwatch being lost or stolen? Why?
 \\[1ex]
8 & How privacy-conscious do you generally consider yourself to be? Why?
 \\[1ex]
9 & How do you think your smartwatch's data could be accessed by companies or other people?
 \\[1ex]
10 & Imagine your smartwatch settings were changed back to their defaults. If you wanted to, what could you do to protect your smartwatch's data? Why?
 \\[1ex]
11 & Imagine your smartwatch settings were changed back to their defaults. If you wanted to prevent apps from tracking your location, what could you do? Why?
 \\[1ex] 
12 & Imagine your smartwatch settings were changed back to their defaults. If you wanted to stop apps from reading your personal data, what could you do? Why?
 \\[1ex]
13 & Imagine your smartwatch settings were changed back to their defaults. If you wanted to limit watch access in case of loss or theft, what could you do? Why?
 \\[1ex]
14 & Could you please show me, and explain aloud, how to disable GPS on your smartwatch?
 \\[1ex]
15 & Could you please show me, and explain aloud, how to change the permissions for a smartwatch app?
 \\[1ex]
16 & Could you please show me, and explain aloud, how to enable a screen lock on your smartwatch?
  \\\midrule
 \# & Protection Motivation Theory (PMT) Questions \\
 \midrule
17 & On a scale from 1 (low) to 10 (high), how serious do you feel the action of your smartwatch data being accessed by a company is? Why?
 \\[1ex] 
18 & How effective do you think smartwatch settings can be in protecting your device's data? Why?
 \\[1ex] 
19 & How able do you feel you are to protect your smartwatch's data? Why?
 \\[1ex]
20 & Do you feel you receive benefits from using data-accessing apps? If so, what?
 \\[1ex]
21 & How much effort do you feel it is to protect your smartwatch's data? Why?
   \\\midrule
 \# & Privacy Paradox Questions \\
 \midrule
22 & We have discussed the use of tools which protect your smartwatch's privacy. Can you think of any techniques or circumstances that would lead you to use these tools more often?
 \\[1ex]
23 & Most of us claim to be concerned about our privacy. However, most of us also fail to fully protect ourselves. This contrast is known as the Privacy Paradox. Why do you think this situation might occur?
 \\[1ex]
24 & You have indicated that you are concerned about your smartwatch's data being accessed. However, on occasions, you didn't use settings to protect that data. Why do you feel this was the case?
 \\
\bottomrule
\end{tabular}
\centering
\label{tbl:interview}
\end{table}

\begin{table}[h!]
\caption{Game Evaluation Questionnaire}
\footnotesize
\setlength{\tabcolsep}{.25em}
\begin{tabular}{m{0.5cm} p{7.3cm}} 
\toprule
\# & Indicate your agreement or disagreement with the statements. \\
& Strongly Agree, Agree, Neutral, Disagree, Strongly Disagree.\\ 
\midrule
1 & ``I found the smartwatch game to be enjoyable.''
 \\[0.5ex]
2 & ``I found the smartwatch game to be usable.''
 \\[0.5ex]
3 & ``I found the smartwatch game to be educational.''
 \\[0.5ex]
4 & ``I found the challenges in the smartwatch game to be easy.''
 \\\midrule
 \# & Qualitative Opinions \\
 \midrule
1 & What did you like most about the smartwatch game? Why?
 \\[0.5ex]
2 & What did you like least about the smartwatch game? Why?
 \\[0.5ex]
3 & What about the game would you like to see improved? Why?
 \\[0.5ex] 
\bottomrule
\end{tabular}
\centering
\label{tbl:evalquestionnaire}
\end{table}

\end{document}